\documentclass[aps,a4paper,twocolumn,10pt,pre,showpacs]{revtex4-2}

\usepackage{graphicx}% Include figure files
\usepackage{dcolumn}% Align table columns on decimal point
\usepackage{bm}% bold math

\usepackage[utf8]{inputenc}
\usepackage[T1]{fontenc}
\usepackage[english]{babel}
\usepackage{amsmath,amssymb}
\usepackage{textcomp}
\usepackage{color}
\usepackage[separate-uncertainty=true]{siunitx}
\usepackage[dvipsnames]{xcolor}
\usepackage[colorlinks=true,allcolors=blue]{hyperref}
\usepackage[normalem]{ulem}
\usepackage{float}
\usepackage{lmodern}% http://ctan.org/pkg/lm

\DeclareSIUnit[number-unit-product = {}]{\bact}{bact}
\graphicspath{{./Figures/}}

\newcommand{\com}[1]{{#1}}

\begin{document}

\title{Direct measurement of the aerotactic response in a bacterial suspension}

\author{J. Bouvard,$^1$  C. Douarche,$^1$ P. Mergaert,$^2$ H. Auradou,$^1$ and F. Moisy$^1$}
\affiliation{%
$^1$Universit\'{e} Paris-Saclay, CNRS, FAST, 91405, Orsay, France.\\
$^2$Université Paris-Saclay, CEA, CNRS, Institute for Integrative Biology of the Cell (I2BC), 91198, Gif-sur-Yvette, France.}

\date{\today}

\begin{abstract}
Aerotaxis is the ability of motile cells to navigate toward oxygen. A key question is the dependence of the aerotactic velocity with the local oxygen concentration $c$. Here we combine simultaneous bacteria tracking and local oxygen concentration measurements using Ruthenium encapsulated in micelles to characterize the aerotactic response of \emph{Burkholderia contaminans}, a motile bacterium ubiquitous in the environment.  In our experiments, an oxygen gradient is produced by the bacterial respiration in a sealed glass capillary permeable to oxygen at one end, producing a bacterial band traveling toward the oxygen source. We compute the aerotactic response $\chi(c)$ both at the population scale, from the drift velocity in the bacterial band, and at the bacterial scale, from the angular modulation of the run times. Both methods are consistent with a power-law $\chi \propto c^{-2}$, in good agreement with existing models based on the biochemistry of bacterial membrane receptors.
\end{abstract}

\maketitle

\section{Introduction}

Many motile bacteria perform chemotaxis: they bias their swimming in response to a gradient of chemicals to move toward an optimum environment~\cite{engelmann1881,pfeffer1887,alicohen1890,beyerinck1893,jennings1901studies,adler1966chemotaxis,adler1966effect,Taylor1999,berg1972chemotaxis,brown1974temporal,dahlquist1972quantitative,ford1991measurement2,lewus2001quantification,gotz1982motility,gotz1987rhizobium}. Aerotaxis~\cite{engelmann1881,Baracchini1959}, the specific response to oxygen, is a common feature of aerobic bacteria~\cite{mazzag2003model,Stricker2020,Frankel1997,popp2014polarity,Fischer2006,
menolascina2017logarithmic,Adler2012,morse2016,Taylor1983}. It was first highlighted by Engelmann~\cite{engelmann1881} in 1881 when he used bacteria to spot sites of oxygen produced by photosynthetic algae.

Aerotaxis plays a key role in many biological processes such as biofilm formation~\cite{Armitano2013,Ardre2015,Holscher2015,Wu2019}, or in the bacterial distribution in the rhizosphere, the soil environment that is under the direct influence of plant roots, where the respiration and the metabolic activities of the roots % of the indigenous microflora 
create oxygen gradients% to which soil bacteria must respond
~\cite{alexandre_2004}. In the field of health, aerotaxis was recently suspected to influence the colonization of epithelial surfaces~\cite{Tamar2016}. This ability to respond positively or negatively to oxygen gradients also opened the possibility to use bacteria to target hypoxic tumours that are resistant to drugs~\cite{felfoul_magneto-aerotactic_2016,baban_bacteria_2010}. It was observed that not all bacteria swim toward a high level of oxygen like the facultative anaerobe \emph{Escherichia coli}~\cite{Adler2012} or other strictly aerobic bacteria~\cite{morse2016,menolascina2017logarithmic}. Indeed, some strains avoid microenvironments too rich in oxygen~\cite{Taylor1983}, while others are microaerophilic and prefer a low but finite level of oxygen~\cite{Frankel1997,popp2014polarity,mazzag2003model,Fischer2006,Stricker2020}. A good qualitative description of aerotaxis is also important to prevent the quality deterioration of food~\cite{Shirai2017}, or to quantify bacterial flux toward regions contaminated by heavy metals~\cite{Stricker2020,Wu2019,Armitano2013}.

%In this paper, we address this issue using an original setup involving the simultaneous measurements of oxygen and bacterial density fluctuations over time.
 %\sout{~\cite{Adler2012} ~\cite{morse2016} 
%This was confirmed experimentally using microfluidic cells where gradients of oxygen can be generated. Combined with video microscopy and image analysis, this method allows to measure $\beta$ in different oxygen environment.This modulation of flagella activity induces a drift velocity $\mathbf{v_A}$ that allows bacteria to migrate along oxygen gradients. Assuming weak changes in local concentration~\cite{keller1971model}, the response to a temporal change in oxygen concentration sensed by a cell along its trajectory can be expressed in terms of spatial gradient, yielding an aerotactic velocity in the form $\mathbf{v_A} = \chi(c) \boldsymbol{\nabla} c$, with $\chi(c)$ the aerotactic sensitivity coefficient. 
%\sout{A key question is the dependence of this coefficient with the oxygen concentration $c$. Larger $\chi$ for decreasing $c$ represents a clear evolutionary advantage, as it yields an enhanced aerotactic response in a strongly depleted environment.} 

Even if the aerotactic behavior of bacteria has been widely studied, the exact dependence of the aerotactic response with the environmental conditions is still an open question.
The  cellular basis of aerotaxis in swimming bacteria is the modulation of their flagellar activity in response to changes in the oxygen level sensed by their receptors~\cite{Hou2000,Stock1997,rebbapragada1997aer,bibikov1997signal,yu_aerotactic_2002}. This translates into a net aerotactic velocity $\bm{v_a}$ which writes, in the limit of small oxygen gradient $\boldsymbol{\nabla} c$, 
\begin{equation}
\label{eq:va}
\bm{v_a} = \chi(c) \boldsymbol{\nabla} c,
\end{equation}
where $\chi(c)$ is the aerotactic response. As a result, aerotactic bacteria that swim toward oxygen-rich regions are able to collectively migrate, forming a traveling band at the population scale.

Traveling bands have been widely investigated for various chemoattractants~\cite{sherris_influence_1957,Baracchini1959,adler1966chemotaxis,adler1966effect,mazzag2003model,saragosti2011directional,cremer2019chemotaxis,Stricker2020}. The discussion is restricted here to aerotaxis, but the concepts are applicable to other chemotaxis. Their dynamics is described by the Keller-Segel equations for the bacterial concentration $b({\bf x},t)$ and oxygen concentration $c({\bf x},t)$, that write in 1D~\cite{keller1971traveling}
\begin{eqnarray}
\dfrac{\partial b}{\partial t} + \dfrac{\partial J_x}{\partial x} =0 ,  \qquad J_x=b v_a - \mu(c) \dfrac{\partial b}{\partial x}, \\
\dfrac{\partial c}{\partial t} = -k b + D_c \dfrac{\partial^2 c}{\partial x^2},
\label{eq:ks}
\end{eqnarray}
with $J_x$ the bacterial flux. The first contribution in $J_x$ is the aerotactic flux, with $v_a = \chi(c) \partial c/\partial x$ the 1D form of the aerotactic velocity \eqref{eq:va}, and the second contribution is the diffusive flux, with $\mu(c)$ the diffusion coefficient due to random swimming (also called random motility coefficient~\cite{rivero1989transport,ford1991measurement2,lewus2001quantification}). The second equation describes the dynamics of the oxygen concentration, with $k$ the oxygen consumption rate by bacteria and $D_c$ the oxygen diffusivity.

The dynamics of the bacterial migration critically depends on the dependence of $\chi$ with $c$. A minimum requirement for the formation of such a traveling band is a decrease of $\chi(c)$ as $c^{-1}$ or steeper~\cite{keller1971traveling}. Models based on the biochemistry of bacterial membrane receptors predict an aerotactic response in the form~\cite{lapidus1976model,rivero1989transport}
\begin{equation}
\label{eq:Chi_Kd_c2}
\chi(c) \propto \dfrac{K_d}{(K_d+c)^2},
\end{equation}
with $K_d$ the dissociation constant of the considered chemoattractant from its receptor. \com{It corresponds to the chemoattractant concentration for which half of the membrane receptors are bound with chemoattractant molecules.} \com{For oxygen, $K_d$ is much smaller than the oxygen content in water at room temperature~\cite{shioi1987oxygen}, suggesting that, except in strongly hypoxic conditions, an effective scaling $\chi \propto c^{-2}$ should be observed.}

\com{Most chemotactic experiments are performed at the population scale, by monitoring the dynamics of the bacterial concentration in a chemical gradient~\cite{adler1966chemotaxis,adler1966effect,Stricker2020,saragosti2011directional}. Experiments at the bacterial scale, initiated by the pioneering work of Brown and Berg~\cite{berg1972chemotaxis}, are more difficult to achieve and are less common in the literature~\cite{brown1974temporal,saragosti2011directional}.}

Inferring the aerotactic response $\chi(c)$ from the dynamics of the bacterial concentration profile evolving in a 1D oxygen gradient is possible in principle, provided that the local oxygen concentration $c(x)$ and $\mu(c)$ are known. Recently, microfluidic setups where the oxygen gradient is fixed by imposing the concentration at the boundaries were developed~\cite{kalinin2009,zhuang2014analytical,kirkegaard_2016,menolascina2017logarithmic,ahmed2010microfluidics}. In this configuration, if the variations of $\mu$ with the oxygen concentration can be neglected, the steady solution of Eq.~\eqref{eq:ks} is an exponential profile $b(x)$ resulting from the balance between the diffusive and the aerotactic fluxes, from which $\chi$ can be obtained. However, accurate measurements can only be obtained for decay lengths falling between a few micrometers up to the channel size of a fraction of millimetres, limiting in turn the range of parameters that can be tested. These studies tend to show the existence of a "logarithmic-sensing" regime, $\chi(c) \propto c^{-1}$, for moderate oxygen gradients (the corresponding aerotactic velocity is $\partial(\ln c)/\partial x$), although other scalings like Eq.~\eqref{eq:Chi_Kd_c2} cannot be excluded over wider range of oxygen gradients. \com{This "log-sensing" was also found in experiments on propagating chemotactic fronts in porous media~\cite{bhattacharjee2021chemotactic,alert2022cellular}.}

The recent development of O$_2$ biocompatible sensors offers now the possibility to perform simultaneous measurements of oxygen concentration and bacterial velocity statistics. In this study, we use Ru-micelles, which consist of Ruthenium (an oxygen-quenched fluorophore) encapsulated in micelles~\cite{douarche2009coli,Adler2012}, to measure the oxygen concentration. \com{Experiments are performed with \emph{Burkholderia contaminans}, a motile bacterium ubiquitous in the environment and present in living organisms. A suspension of \emph{B. contaminans} of various concentrations mixed with Ru-micelles is enclosed in an elongated glass capillary closed at one end but permeable to air on the other, allowing an unidirectional oxygen gradient.} The bacterial respiration lowers the oxygen concentration in the suspension, which in turn strengthens the gradient, resulting in a traveling band and a strong accumulation of bacteria near the oxygen source at large time. The simultaneous measurements of oxygen and bacterial tracks allow us to compute directly the aerotactic response $\chi(c)$, both at the population scale and at the bacterial scale. Our results suggest a robust $\chi(c) \propto c^{-2}$ scaling, consistent with Eq.~\eqref{eq:Chi_Kd_c2} for a dissociation constant $K_d$ much smaller than the ambient oxygen concentration level.

\newpage % à mettre/enlever en fonction de la mise en page
\section{Material and methods}

\subsection{Experimental setup}
\label{sec:setup}

%%%%%%%%%%%%%%%%%%%%%%%%%%%
\begin{figure}[b]
\includegraphics[trim = 0mm 0mm 0mm 0mm, clip, width=0.48\textwidth, angle=0]{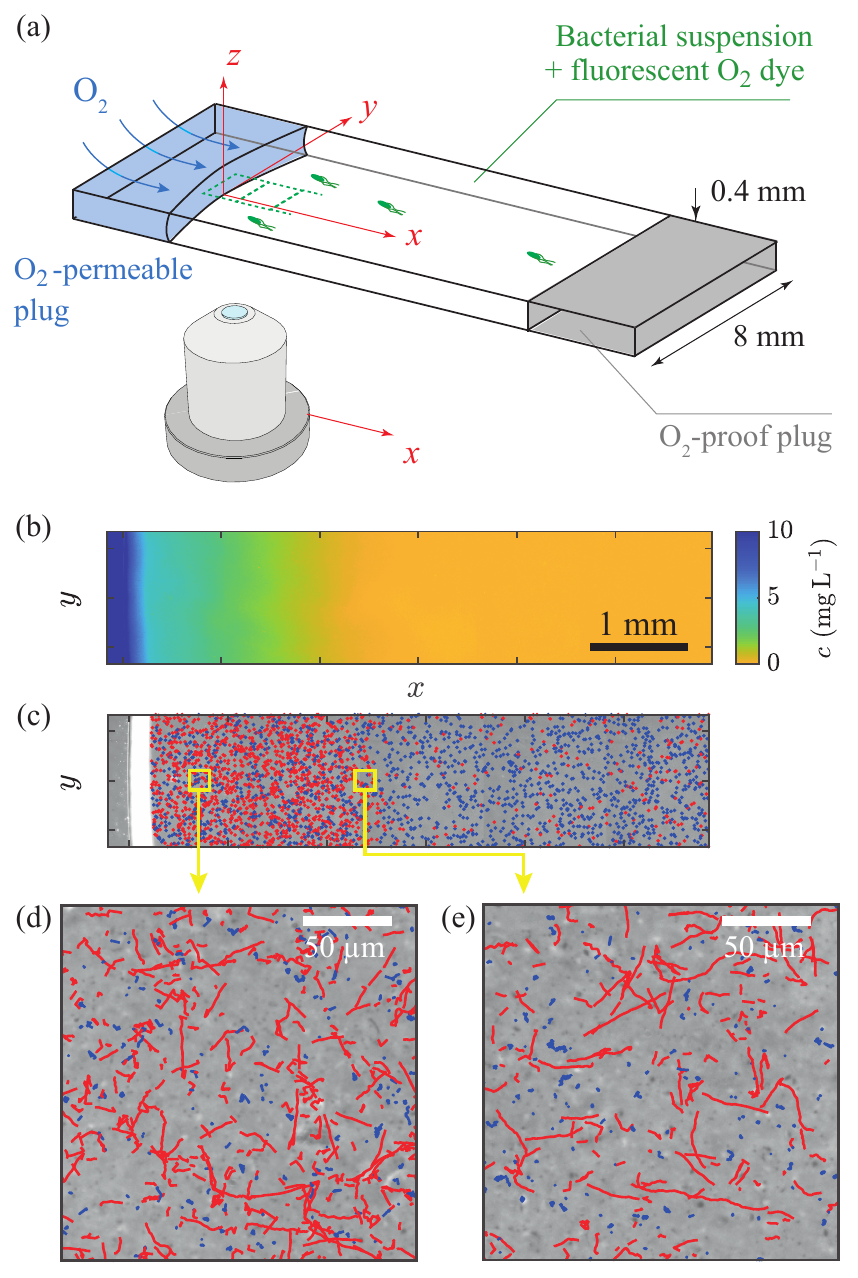} 
\caption{Aerotactic migration band in a glass capillary permeable to air on one side. (a) Experimental setup, made of a glass capillary of rectangular section, filled with the mixture of bacterial suspension and fluorescent oxygen dye. One side of the capillary is sealed with an oxygen-porous PDMS plug, while the other side is sealed with an oxygen-proof grease plug. \com{(b) Oxygen concentration field during the migration of the traveling band, computed from the fluorescence light intensity}. (c) Images of bacteria at the same time, superimposed to the bacterial tracks (nonmotile and motile bacteria are in blue and red, respectively). (d), (e) Magnification at two locations, showing the nearly random tracks at large oxygen concentration and the strongly biased tracks at low oxygen concentration.}
\label{fig:setup}
\end{figure}
%%%%%%%%%%%%%%%%%%%%%%%%%%%%%

The experimental setup is shown in  Fig.~\ref{fig:setup}(a). A homogeneous suspension of \emph{B. contaminans} bacteria is enclosed in a glass capillary (Vitrotubes) of rectangular cross section (\SI{0.4}{\mm} height, \SI{8}{\mm} width) and \SI{50}{\mm} long. The capillary is closed at one end by PDMS (polydimethylsiloxane) plug, a biocompatible cured polymer permeable to oxygen, and sealed at the other end by grease \com{(vaseline Cooper)}. The PDMS plug is made by dipping the capillary in liquid PDMS, resulting in a plug of $13\pm 0.5$\,mm long.

\com{The bacterial strain of \emph{B. contaminans} used in this study is an environmental strain characterized from the sequencing of \emph{16S} and \emph{recA} gene fragments. It was selected for its strong aerotactic ability in a preliminary screening of available strains.} Bacteria are grown overnight in YEB medium (yeast extract beef: \SI{5}{\g\per\liter} beef extract, \SI{1}{\g\per\liter} yeast extract, \SI{5}{\g\per\liter} peptone, \SI{5}{\g\per\liter} sucrose, \SI{0.5}{\g\per\liter} MgCl$_2$) at \SI{28}{\celsius} in an orbital shaker running at \SI{200}{rpm} up to an optical density $\text{OD}_{600}=5.5\pm0.5$. The suspensions are then diluted in YEB to the desired concentration, and \SI{1}{\umol\per\liter} of Ru-micelles~\cite{douarche2009coli} is added to the suspension. The solution is finally homogenised with an orbital vortex mixer before injection in the capillary. Experiments were repeated for three bacterial concentrations, denoted hereafter by their optical densities $\text{OD}=0.05$, 0.1 and 0.2 ($\SI{1}{OD} \sim \SI{1.8e6}{\bact\per\uL}$).

The capillary is located on the stage of an inverted microscope (Leica DMI4000B) equipped with a long working distance 10X objective and a Hamamatsu Orca Flash 4 camera allowing to record fields of view of $1.33 \times 1.33$\,mm$^2$. Two imaging methods are used sequentially: fluorescence imaging, from which the oxygen concentration is computed (see Fig.~\ref{fig:setup}(b) and Appendix~\ref{sec:SM_O2}), and phase contrast for bacteria tracking [Fig.~\ref{fig:setup}(c)]. \com{Measurements are performed in a plane at midheight in the channel, with a depth of field of \SI{30}{\um}, thereby selecting bacteria swimming far from the walls}. Measurements are performed at 5 locations along the $x$ axis with \SI{10}{\%} overlap, covering a distance of nearly \SI{6}{\mm} in the $x$-direction from the source at $x=0$.

\com{The migration band travels along the capillary tube until it reaches a stationary state, which consists in a strong accumulation of motile bacteria in a narrow region near the oxygen source. The timescale of the migration is approximately \SI{100}{min} for the lowest bacterial concentration ($\mathrm{OD}=0.05$) and \SI{32}{min} for the highest ($\mathrm{OD}=0.2$). These times are smaller than the doubling time of this strain, $T \simeq 170$\,min, measured in the exponential growth phase in an oxygenated culture medium, so the bacterial growth can be neglected in our experiments.}

\com{Images are taken during the band migration using the following acquisition sequence:} a fluorescence picture is first taken at each of the five locations, then a movie of 100 images at \SI{20}{Hz} is acquired in phase contrast at the same locations. The sequence is repeated up to 20 times during the bacterial migration, with a time step between each scan of 3 to 10 minutes depending on the time scale of the band migration. The maximum time lag between an oxygen measurement at a given location $x$ and the bacterial tracks at the same location is less than \SI{90}{\s}, a value comfortably smaller than the migration timescale.

\subsection{Bacteria tracking}
\label{sec:bact_track}

\begin{figure*}
\includegraphics[trim = 0mm 0mm 0mm 0mm, clip, width=0.99\textwidth, angle=0]{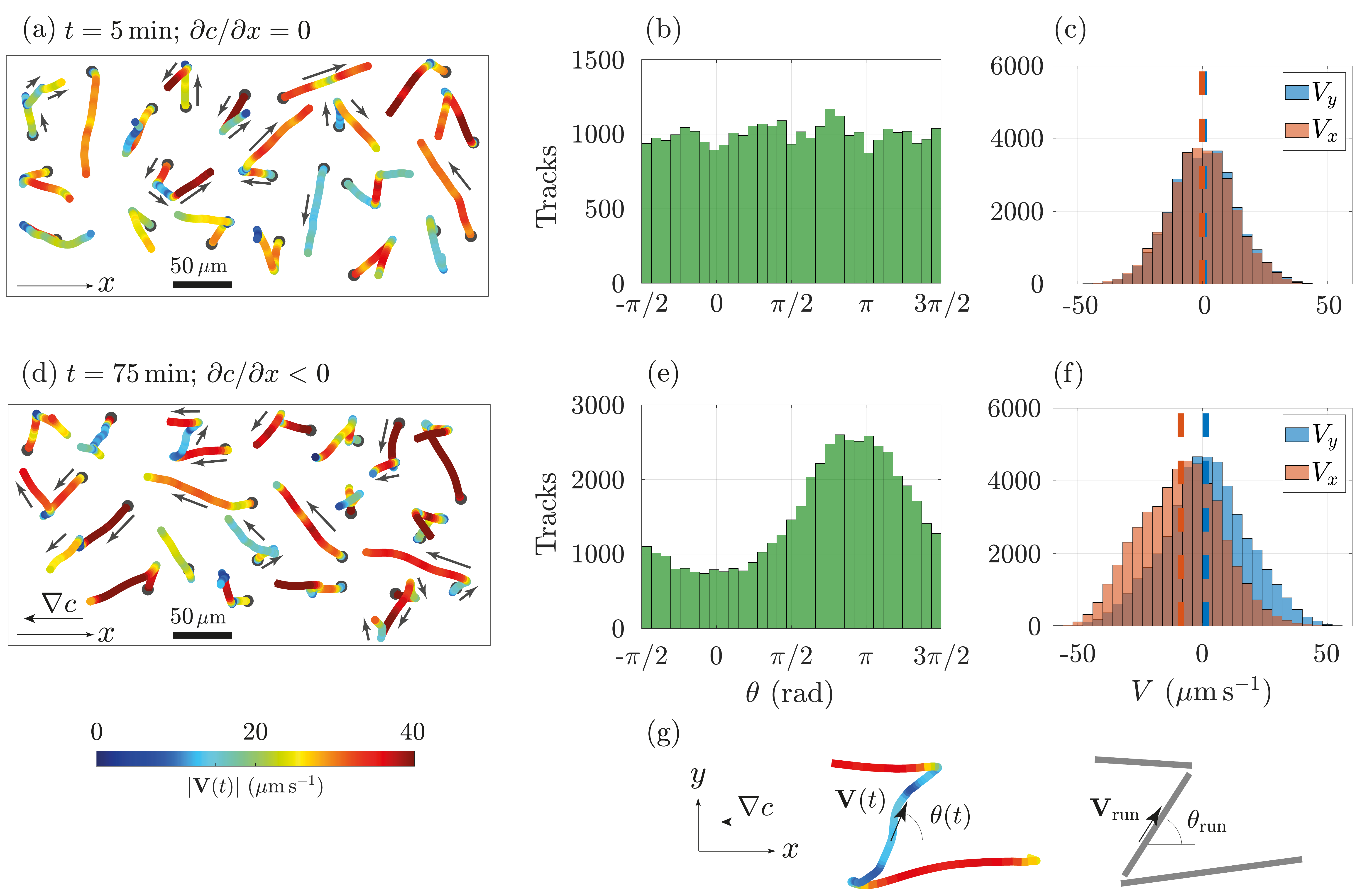} 
\caption{Characteristics of bacterial tracks without and with aerotaxis. Data from the experiment at $\mathrm{OD}=0.05$ \com{($\sim \SI{9e4}{\bact\per\uL}$)}, at location $x=\SI{4500\pm500}{\um}$ from the oxygen source. (a), (b), (c) $t=\SI{5}{\min}$, while the oxygen concentration is homogeneous in the glass capillary; (d), (e), (f) $t=\SI{75}{\min}$, when a steep oxygen gradient is formed. (a), (d) Selection of bacterial tracks, coloured according to their instantaneous velocity. The starting point of a track is represented with a grey circle. The track durations are between 3 and \SI{5}{\s}. (b), (e) Histograms of the instantaneous trajectory angle $\theta$. (c), (f) Histograms of the instantaneous velocity components $V_x,V_y$. (g) Segmentation of tracks into a set of runs, defining the run velocity ${\bf V}_\mathrm{run}$ and run angle $\theta_\mathrm{run}$ with respect to $x$.}
\label{fig:tracks}
\end{figure*}

From the sequences of phase contrast images, the bacterial tracks are computed using Trackmate~\cite{tinevez2017trackmate} and post-processed using an in-house Matlab code. For each track $i$, we measure the two-dimensional positions ${\bf X}_i(t)=(X_i(t),Y_i(t))$, from which we compute the two-dimensional velocities ${\bf V}_i(t) = \delta {\bf X}_i / \delta t$ using a sampling time $\delta t$. This sampling time is chosen such as to separate the motile and nonmotile populations that are naturally present in the suspension in approximately the same proportions [red and blue tracks in Fig.~\ref{fig:setup}(d) and (e)]. During the time $\delta t$, motile cells travel a distance $\simeq V_0\,\delta t$, with $V_0 \simeq \SI{20}{\um\per\s}$ the typical swimming velocity (see Fig.~\ref{fig:tracks}). During the same time $\delta t$, nonmotile cells move randomly over a distance $\simeq \sqrt{D \, \delta t}$, with $D \simeq \SI{0.1}{\square\um\per\s}$  the thermal diffusion coefficient (as measured independently from the mean-square displacement~\cite{bouvard2022thesis}). A clear separation between the two populations is obtained by using a time interval $\delta t \simeq 10^3 \, D / V_0^2  \simeq \SI{200}{\ms}$. With this value, nonmotile bacteria have an effective velocity $\sqrt{D / \delta t} \simeq \SI{1}{\um\per\s}$ much smaller than that of motile bacteria, and can be filtered out by applying a velocity threshold at $\SI{6}{\um\per\s}$. Only motile bacteria are considered in the following.

%%%%%%%%%%%%%%%%%%%%%%%%%%%
%%%%%%%%%%%%%%%%%%%%%%%%%%%%%

A selection of tracks is shown in Fig.~\ref{fig:tracks}, without [Fig.~\ref{fig:tracks}(a)] and with [Fig.~\ref{fig:tracks}(d)] aerotaxis. These tracks display a succession of approximately straight runs, typically 10 to \SI{100}{\um}, separated by rapid changes in orientation. A specificity of the swimming pattern of \emph{B. contaminans} is the frequent occurrence of very sharp turning angles close to \SI{180}{\degree} (see Appendix~\ref{sec:SM_run-reverse}), contrarily to the smaller mean turning angle of $\simeq \SI{70}{\degree}$ of the classical "run-and-tumble" pattern in the model bacterium \emph{E. coli}~\cite{berg1972chemotaxis,berg1993random}. Such a "run-reverse" pattern is also encountered in some marine bacteria~\cite{barbara2003bacterial,taktikos2013motility,grognot2021multiscale}. In the presence of the oxygen gradient, runs in the direction of the oxygen source ($x<0$) are longer and more frequent. This bias is difficult to see directly in Fig.~\ref{fig:tracks}(a) and \ref{fig:tracks}(d), but appears clearly in the distributions of velocity components and angle in Fig.~\ref{fig:tracks}(b), \ref{fig:tracks}(c), \ref{fig:tracks}(e), \ref{fig:tracks}(f). The instantaneous swimming angle $\theta(t)$ is defined in  Fig.~\ref{fig:tracks}(g), with $\theta=\pi$ toward the oxygen source ($x<0$). In the presence of the oxygen gradient, the distribution of angles shows a strong asymmetry, with tracks in the oxygen direction approximately three times more frequent than in the opposite direction. Interestingly, the distribution of $V_x$ is peaked to a negative value, but its width is comparable to that of $V_y$. This feature is robust at all times, with a ratio of standard deviations $\sigma(V_x) / \sigma(V_y) \simeq 1.00 \pm 0.01$, indicating that the bacterial motion can be approximately modelled as the sum of an isotropic diffusion and a small deterministic drift toward $x<0$.

We determined the diffusivity of the bacteria without aerotaxis in two setups: in a chamber fully permeable to oxygen (with a PDMS cover) and in a fully sealed capillary (capillary glass sealed with a grease plug at both ends). In the first setup, the oxygen concentration remains at saturation level at all time, while in the second it linearly decreases in time because of bacterial respiration. From the bacterial tracks, we compute the diffusivity as $\mu = (1/2) V_c^2 T$, where $V_c$ and $T$ are determined from an exponential fit of the two-dimensional velocity correlation function (see Appendix~\ref{sec:SM_mu}). We obtain $\mu \simeq \SI{450\pm100}{\square\um\per\s}$, with no significant evolution in time in both setups as long as oxygen is available. In the sealed capillary, the motility rapidly drops when oxygen is exhausted and transitions to a Brownian diffusive motion. Similarly, the mean 2D velocity stays constant throughout both experiments, and only drops when the oxygen concentration reaches zero in the sealed capillary experiment.

\section{Dynamics of the aerotactic front}
\label{sec:macro}

To quantify the dynamics of the aerotactic migration of bacteria toward the oxygen source, we compute the velocity statistics conditioned on the $x$ position along the capillary. For this, we first define for each track $i$ its average position $\overline{{\bf X}}_i=(\overline{X}_i,\overline{Y_i})$ and average velocity $\overline{\bf V}_i$. Then, we average the velocity norm and $x$-component conditioned on the averaged $x$-position,
\begin{eqnarray}
    v(x,t) &=& \langle | \overline{\bf V}_i | ; \, x = \overline{X}_i \rangle, \label{eq:vcond} \\
    v_x(x,t) &=& \langle  \overline{\bf V}_i \cdot {\bf e}_x ; \, x = \overline{X}_i \rangle.
\label{eq:vxcond}
\end{eqnarray}
The brackets $\langle \cdot \rangle$ denote the average over all tracks $i$ satisfying this spatial conditioning. In practice, the conditioning is performed on a window size of width $\Delta x = \SI{200}{\um}$. This window size is chosen larger than the typical run length, but smaller than the characteristic length scale of the oxygen and bacterial concentration profiles. Note that the velocities in Eqs.~\eqref{eq:vcond} and \eqref{eq:vxcond} are the two-dimensional projections of the true bacteria motion. This implies that for bacteria swimming isotropically at constant (three-dimensional) velocity $V_0$, we have $v=\pi V_0/4 \simeq 0.78 V_0$, whereas for bacteria swimming in the microscope plane we simply have $v=V_0$.

%%%%%%%%%%%%%%%%%%%%%%%%%
\begin{figure}[H]
\centering
\includegraphics[trim = 0mm 0mm 0mm 0mm, clip, width=0.38\textwidth, angle=0]{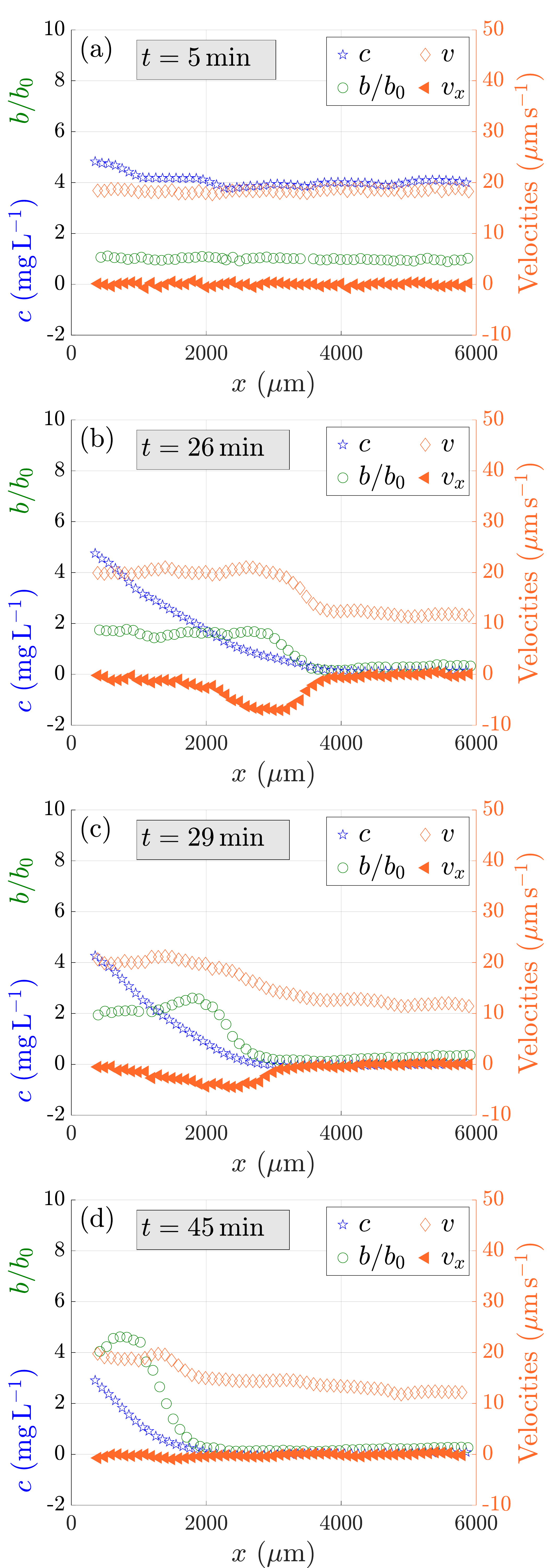} 
\caption{Variations of the local density of bacteria $b(x)$ (green circles), oxygen concentration $c(x)$ (blue stars), magnitude of the swimming velocity of bacteria $v(x)$ (empty orange diamonds) and drift velocity along the gradient $v_x(x)$ (filled orange diamonds) as functions of the distance from the oxygen source at four different times during the migration of the bacterial band. The capillary is initially filled with an homogeneous suspension of bacteria at a concentration \com{$\text{OD}=0.2$} \com{($\sim \SI{3.6e5}{\bact\per\uL}$)}. Time is counted from the sealing of the capillary.}   
\label{fig:figure2}
\end{figure}
%%%%%%%%%%%%%%%%%%%%%%%%%

The dynamics of the migration band is depicted in Fig.~\ref{fig:figure2} for \com{$\text{OD}=0.2$} \com{($\sim \SI{3.6e5}{\bact\per\uL}$)}, along with the oxygen concentration profile, the mean velocity norm $v(x)$ and the drift velocity along the gradient $v_x(x)$. Five minutes after the sealing of the capillary [Fig.~\ref{fig:figure2}(a)], the oxygen level has slightly decreased but remains essentially homogeneous, and no measurable drift velocity nor variations in the bacterial concentration are detected. \com{At $t=\SI{26}{\min}$ [Fig.~\ref{fig:figure2}(b)], the migration band has reached the middle of the measurement area, with the back end of the band ($x\simeq\SI{3.5}{\mm}$) corresponding to the location where the oxygen level is zero. In the migration band, the drift velocity reaches $v_x \simeq \com{\SI{-7}{\um\per\s}}$ (negative velocity is toward the oxygen source), which is \SI{30}{\%} of the 2D mean swimming velocity.} \com{Three} minutes later [Fig.~\ref{fig:figure2}(c)], the bacterial band is more pronounced as it approaches the oxygen source, but it travels at a smaller velocity, $v_x \simeq \com{\SI{-4}{\um\per\s}}$. After \com{\SI{45}{\min}}, an asymptotic state is reached, with a strong accumulation of bacteria near the $\text{O}_2$ source and $v_x \simeq \SI{0}{\um\per\s}$ [Fig.~\ref{fig:figure2}(d)].

\begin{figure}[b]
\centering
\includegraphics[trim = 2mm 65mm 4mm 70mm, clip, width=0.48\textwidth, angle=0]{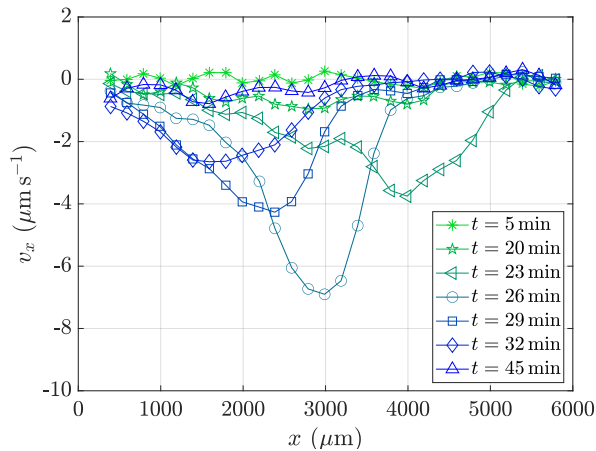}
\caption{Drift velocity \com{$v_x(x)$} as a function of the distance from the oxygen source at different times during the migration of the bacterial band. The initial bacterial concentration is $\text{OD}=0.2$.}   
\label{fig:Vx_x}
\end{figure}

\com{The temporal evolution of the drift velocity $v_x(x)$ is summarized at all times in Fig.~\ref{fig:Vx_x}, showing a net migration band toward the oxygen source as time proceeds. The experiment was repeated with the two lower initial bacterial concentrations, OD = 0.1 and 0.05 (see Appendix~\ref{sec:SM_compil_profiles}), yielding a similar dynamics but on a timescale that varies approximately as the inverse of the bacterial concentration.}

%%%%%%%%%%%%%%%%%%%%%%%%%
%%%%%%%%%%%%%%%%%%%%%%%%%

We now turn to the determination of the aerotactic coefficient $\chi(c) = v_a / (\partial c/\partial x)$, where $v_a$ is the aerotactic velocity.  A difficulty in inferring $v_a$ from the measured drift velocity $v_x$ is that $v_x$ is related to the {\it total} bacterial flux $J_x=bv_x$, and therefore includes both the intrinsic aerotactic contribution $bv_a$ and the diffusive contribution $\mu(c) \partial b/\partial x$ [Eq.~\eqref{eq:ks}]. Correcting for this diffusive contribution is delicate, because the dependence of  $\mu$ with $c$, which stems from the dependence of $\chi$ with $c$, is difficult to infer from the data. We therefore focus on regions where $\partial b / \partial x \simeq 0$, near the maximum of the bacterial density in the migrating band or close to the O$_2$ source. In these regions, the diffusive flux is small and we have $v_x \simeq v_a$, so the aerotactic response is simply given by
\begin{equation}
    \chi_\mathrm{macro}(c) = \frac{v_x}{\partial c/\partial x},
\end{equation}
which can be directly computed from our velocity and oxygen concentration measurements.  We note $\chi_\mathrm{macro}$ this first determination of the aerotactic coefficient, as it is obtained from the drift velocity at the macroscopic level.

%%%%%%%%%%%%%%%%%%%%%%%%%
\begin{figure}[tb]
\includegraphics[trim = 1mm 54mm 8mm 60mm, clip, width=0.48\textwidth, angle=0]{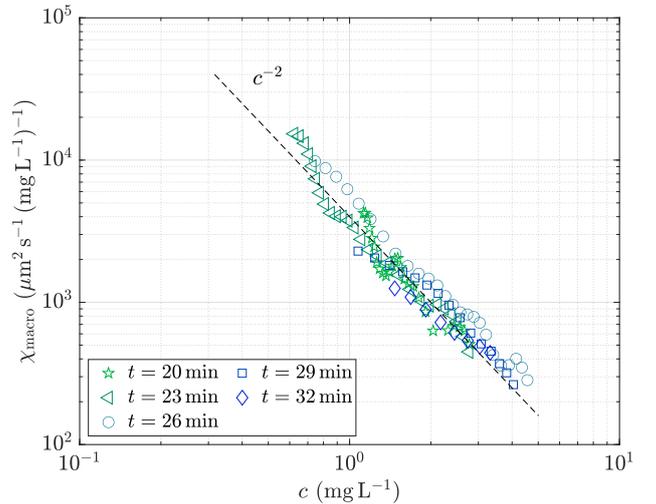}
\caption{Aerotactic response $\chi_{\mathrm{macro}} \simeq v_x/({\partial c}/{\partial x})$  as a function of the local oxygen concentration $c$. The values are computed from the drift velocity $v_x(x)$ and oxygen gradient $\partial c/\partial x$ measured on the maximum of the migration band (where $\partial b/\partial x \simeq 0$), for an initial bacterial concentration $\text{OD}=0.2$.}
\label{fig:figure3}
\end{figure}
%%%%%%%%%%%%%%%%%%%%%%%%%

Figure~\ref{fig:figure3} shows the aerotactic response $\chi_\mathrm{macro}$ as a function of $c$ at different times during the migration. \com{It shows a well-defined power-law decay $\chi_\mathrm{macro} \propto c^{-n}$ in the range $c=\SIrange{0.6}{5}{\mg\per\liter}$, with a best fit exponent $n \simeq 1.9 \pm 0.4$}. The upper bound of this range is limited by the smallest measurable drift velocity, while the lower bound is limited by the resolution of the fluorescence signal. The good collapse of the data at different times on a single curve indicates that the aerotaxis response is governed by the local oxygen concentration alone, with no measurable delay, so it can be considered as approximately instantaneous in our experiment.

\com{This scaling is consistent with $\chi_\mathrm{macro} \propto c^{-2}$, which corresponds to Eq.~\eqref{eq:Chi_Kd_c2} in the limit of a small dissociation constant, $K_d \ll c$.} Although the value of $K_d$ of oxygen is not known for \emph{B. contaminans}, the absence of visible cutoff in $\chi_\mathrm{macro}(c)$ at small $c$ suggests that $K_d$ is much smaller than the minimal oxygen concentration measured in Fig.~\ref{fig:figure3} ($\simeq \SI{0.6}{\mg\per\liter}$). This is consistent with the value of $K_d \simeq \SI{0.02}{\mg\per\liter}$ determined for other bacterial species like \emph{Escherichia coli} and \emph{Salmonella typhimurium}~\cite{shioi1987oxygen}.

%%%%%%%%%%%%%%%%%%%%%%%%%
\begin{figure}[tb]
\includegraphics[trim = 8mm 63mm 19mm 70mm, clip, width=0.48\textwidth, angle=0]{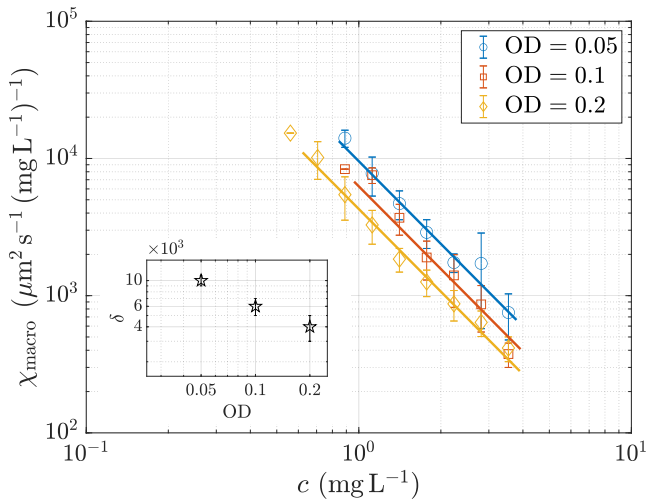}
\caption{Aerotactic response $\chi_{\mathrm{macro}} \simeq v_x/({\partial c}/{\partial x})$ computed on the maximum of the migration band (where $\partial b/\partial x \simeq 0$), plotted as a function of the oxygen concentration $c$ for different initial bacterial concentrations: $\text{OD}=0.05$ (blue circles), $\text{OD}=0.1$ (orange squares) and $\text{OD}=0.2$ (yellow diamonds). The lines show the best fit $\chi_{\mathrm{macro}}=\delta/c^2$ for each OD, the coefficient $\delta$ being plotted as a function of the OD in the inset. For $c>\com{\SI{5}{\mg\per\liter}}$, the aerotactic velocity is too close to zero for a reliable measurement of $\chi_\mathrm{macro}$.}
\label{fig:chi_macro}
\end{figure}
%%%%%%%%%%%%%%%%%%%%%%%%%

We have repeated the measurement of $\chi(c)$  for the three bacterial concentrations. The curves are summarised in Fig.~\ref{fig:chi_macro}, with error bars reflecting the dispersion for each OD. Here again, the data is well described by the power law $\chi \simeq \delta c^{-2}$, but with a coefficient $\delta$ that decreases as the bacterial concentration increases (see inset).  This suggests that the scaling of the aerotactic response with oxygen concentration is robust, but the level of this response also depends on other parameters related to the bacterial concentration. The delay between the measurements of the oxygen concentration and the bacterial tracks cannot explain this dependence. Because of the sequential acquisition procedure (see Sec.~\ref{sec:bact_track}), the oxygen concentration is measured approximately \SI{1}{\min} before the bacterial velocity. During this time lag, $c$ decreases while $|\partial c/\partial x|$ increases at a given $x$, at a rate which increases with the OD; this should increase the apparent $\delta$, which does not match our observations. A second possible bias is a selection of the fastest swimming bacteria as the OD is increased; however, this sorting effect would also yield an increase of the apparent $\delta$, which again is not compatible with our data. We can conclude that the observed decrease of $\delta$ with the OD is a genuine feature of the aerotactic response of this strain, that may result from a secondary negative chemotactic response lowering the effect of the primary positive aerotactic response.

\section{Aerotactic run time modulation}
\label{sec:micro}

Measuring the aerotactic coefficient $\chi(c)$ from the mean drift velocity $v_x$ is inherently limited by the effect of diffusion, which restricts our measurements to regions where the diffusive flux vanishes, for $\partial b/\partial x \simeq 0$. To circumvent this limitation, we now investigate the aerotaxis at the microscopic level, with the aim to characterise the modification of the swimming pattern statistics due the oxygen gradient.

For each track, we define a run as a segment between two sharp reorientations [Fig.~\ref{fig:tracks}(g)], such that \com{$|\theta(t+\delta t) - \theta(t)| > \Delta \theta_c$, with $\Delta \theta_c$ a threshold. For the sampling time used here, $\delta t = \SI{50}{\ms}$, robust results are obtained using the threshold $\Delta \theta_c = 0.8$\,rad.}  \com{The reorientation time between two runs is of the order of \SIrange{0.1}{0.25}{\s} (the limited sampling frequency does not allow a finer characterisation of its distribution).} For each run $j$, we define the mean velocity $\mathbf{V}_{\mathrm{run},j}$, the mean angle $\theta_{\mathrm{run},j}$ with respect to the $x$ axis and the run time $\tau_{\mathrm{run},j}$.

\begin{figure}[b]
\includegraphics[width=0.47\textwidth]{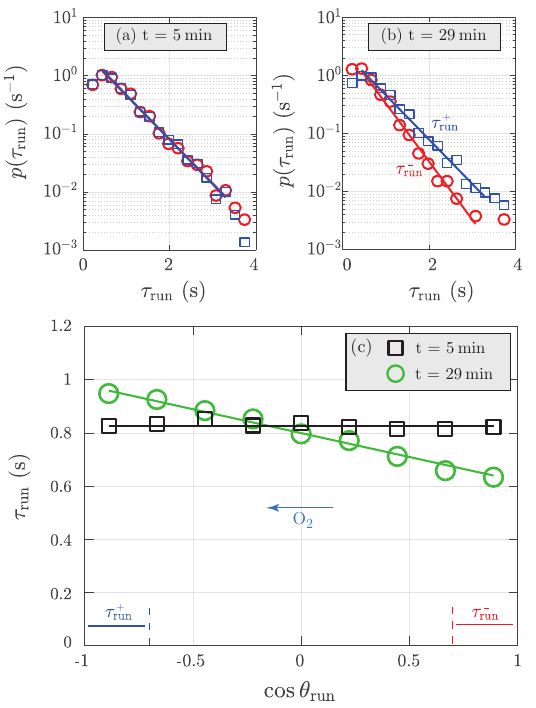} 
\caption{\com{Distributions of run times $\tau_{\mathrm{run}}^\pm$ before the aerotactic band [(a), $t=\SI{5}{\min}$] and in the aerotactic band [(b), $t=\SI{29}{\min}$], for $\mathrm{OD}=0.2$, $x\simeq \SIrange{2.5}{3}{\mm}$. Blue: $\tau_{\mathrm{run}}^+$, conditioned on $ \theta_{\mathrm{run},j} = 180 \pm 45^\mathrm{o}$ (toward the oxygen source); red: $\tau_{\mathrm{run}}^-$, conditioned on $\theta_{\mathrm{run},j} = 0 \pm 45^\mathrm{o}$. The lines show exponential fits $\exp(-\beta \tau_{\mathrm{run}}^{\pm})$, with $\beta^\pm \simeq \SI{1.9}{\per\s}$ in panel (a), and $\beta^- \simeq \SI{2.3}{\per\s}$, $\beta^+ \simeq \SI{1.7}{\per\s}$ in panel (b).
(c) Average run time $\tau_\mathrm{run}$ as a function of $\cos\theta_\mathrm{run}$ before the aerotactic band (black symbols) and in the aerotactic band (green symbols). The green line shows the linear fit using Eq.~\eqref{eq:taurun} with $\alpha \simeq 0.22$. The vertical dashed lines indicate the ranges of run angles $\theta_{\mathrm{run}}$ (such that $| \cos \theta_{\mathrm{run}}| > 1/\sqrt{2}$) over which the run time distributions $p(\tau_{\mathrm{run}}^\pm)$ in panels (a, b) are computed.}}
\label{fig:taurun_distri}
\end{figure}

To characterise the aerotactic bias of the run times, we partition the runs \com{in two sets specified by the swimming direction, toward the oxygen source ($\theta_{\mathrm{run},j} = 180 \pm 45^\mathrm{o}$) and in the opposite direction ($\theta_{\mathrm{run},j} = 0 \pm 45^\mathrm{o}$), and define the corresponding run times as $\tau_{\mathrm{run}}^+$ and $\tau_{\mathrm{run}}^-$, respectively.} Figure~\ref{fig:taurun_distri} shows the distribution of these run times, with and without aerotaxis. At early time [Fig.~\ref{fig:taurun_distri}(a)], without significant oxygen gradient, $\tau_{\mathrm{run}}^+$ and $\tau_{\mathrm{run}}^-$ follow the same Poisson distribution, $p(\tau_\mathrm{run}^{\pm}) \simeq \exp(-\beta \tau_{\mathrm{run}}^{\pm})$, with $\beta \simeq\SI{1.9}{\per\s}$. At larger time [Fig.~\ref{fig:taurun_distri}(b)], in the migration band, the two distributions still decay exponentially but with different decay rates, confirming that bacteria swim longer when moving toward the oxygen source.

%%%%%%%%%%%%%%%%%%%%%%%%%%%%%%%%%
%%%%%%%%%%%%%%%%%%%%%%%%%%%%

To further quantify this effect, we compute, for each time $t$ and location $x$ along the capillary, the mean run duration conditioned on the mean run angle,
\begin{equation}
        \tau_{\mathrm{run}}(\theta_{\mathrm{run}}) = \langle  \tau_{\mathrm{run,}j} ; \, \theta_{\mathrm{run,}j} = \theta_{\mathrm{run}} \rangle,
\end{equation}
with $\langle \cdot \rangle$ the average over the runs $j$.
We plot this mean run time $\tau_{\mathrm{run}}$ as a function of $\cos \theta_{\mathrm{run}}$  in Fig.~\ref{fig:taurun_distri}(c), both before and during the migration band. We observe a well-defined linear trend,
\begin{equation}
     \tau_{\mathrm{run}} (\theta_{\mathrm{run}}) = \tau_0 (1 - \alpha \cos \theta_{\mathrm{run}}),
\label{eq:taurun}
\end{equation}
with $\alpha$ the run time modulation coefficient, and $\tau_0$ the mean run time at $\cos\theta_{\mathrm{run}}=0$. Such a linear dependency of $\tau_{\mathrm{run}} $ with $\cos \theta_{\mathrm{run}}$ comes from the fact that bacteria are sensitive to the time variation of the local oxygen gradient sampled along their trajectory, $d c/d t \simeq {\bf V} \cdot \nabla c = V \cos \theta \, \partial c / \partial x$ (see Appendix~\ref{app:artm}). Without oxygen gradient, we have $\alpha \simeq 0$ (isotropic swimming), whereas in the oxygen gradient, runs toward the oxygen source are $(1+\alpha)$ longer than $\tau_0$ while runs away from the oxygen source are $(1-\alpha)$ shorter than $\tau_0$.

\begin{figure}[bt]
\includegraphics[trim = 12mm 63mm 18mm 70mm, clip, width=0.48\textwidth, angle=0]{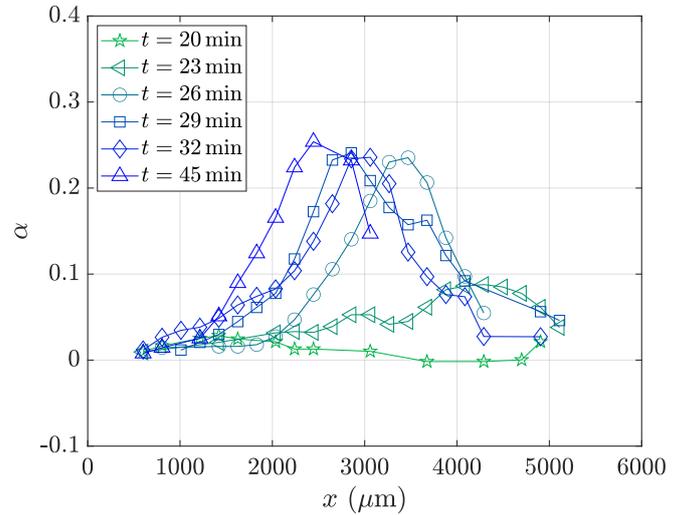} 
\caption{Run time modulation coefficient $\alpha$ as function of the distance from the oxygen source at different times during the migration of the bacterial band. Same data as in Fig.~\ref{fig:Vx_x} \com{($\text{OD}=0.2$)}.}
\label{fig:alpha_x}
\end{figure}

From the run statistics, we extract the run time modulation coefficient $\alpha$ at each time, for each location $x$ along the capillary chamber, and for the three bacterial concentrations ($\mathrm{OD}=0.05$, 0.1 and 0.2). The time evolution of $\alpha(x)$, plotted in Fig.~\ref{fig:alpha_x} for $\mathrm{OD}=0.2$, shows a clear drift of the aerotactic response toward the oxygen source as the migration band proceeds, which is consistent with the measured drift velocity $-v_x(x)$ in Fig.~\ref{fig:Vx_x}. These two quantities can be related if we assume that the swimming velocity $V_0$ and the mean run time $\tau_0$ do not depend on $\theta_{\mathrm{run}}$, yielding (see Appendix~\ref{app:artm})
\begin{equation}
\label{eq:va_micro}
    v_a = -V_0 \dfrac{\alpha}{2}.
\end{equation}
Interestingly, this estimate of $v_a$ from $\alpha$ is expected to hold everywhere, not only at the maximum of the migration band where the diffusion contribution vanishes. This is clearly visible by comparing Figs.~\ref{fig:Vx_x} and \ref{fig:alpha_x}: while $v_x$ drops to 0 at long time because aerotaxis is balanced by diffusion in the asymptotic state, a strong aerotactic response is still present at the bacterial scale in the run time modulation coefficient $\alpha$.

%%%%%%%%%%%%%%%%%%%%%%%%%%%%
%%%%%%%%%%%%%%%%%%%%%%%%%%%%

\begin{figure}[tb]
\includegraphics[trim = 8mm 63mm 19mm 70mm, clip, width=0.48\textwidth, angle=0]{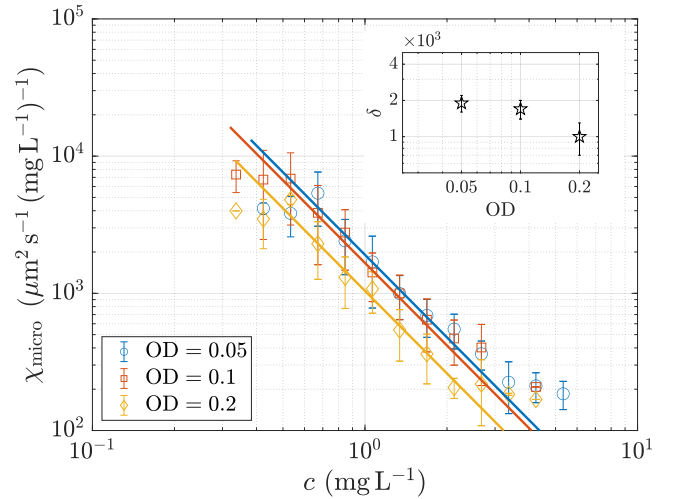}
\caption{Aerotactic response $\chi_{\mathrm{micro}} \simeq -\alpha v/(2{\partial c}/{\partial x})$ plotted as a function of the oxygen concentration $c$ for different bacterial concentrations: $\text{OD}=0.05$ (blue circles), $\text{OD}=0.1$ (orange squares) and $\text{OD}=0.2$ (yellow diamonds). The lines show the best fit $\chi_{\mathrm{micro}}=\delta/c^2$ for each OD, the coefficient $\delta$ being plotted as a function of the OD in the inset.} %For $c>\SI{5}{\mg\per\liter}$, the data is indistinguishable from zero.
\label{fig:chi_micro}
\end{figure}

We finally plot in Fig.~\ref{fig:chi_micro} the aerotactic coefficient [Eq.~\eqref{eq:va}] computed using Eq.~\eqref{eq:va_micro}, which we call $\chi_{\mathrm{micro}}$. This coefficient is again compatible with the scaling $c^{-2}$ for all bacterial concentrations. Since we are not limited here to $\partial b/\partial x \simeq 0$, the range of oxygen concentration where $\chi_{\rm micro}$ can be computed is extended to lower oxygen concentrations, down to $c \simeq \com{\SI{0.3}{\mg\per\liter}}$.

%%%%%%%%%%%%%%%%%%%%%%%%%
%%%%%%%%%%%%%%%%%%%%%%%%%

Although the aerotactic responses computed at the population scale (Fig.~\ref{fig:chi_macro}) and at the bacterial scale (Fig.~\ref{fig:chi_micro}) both display the same scaling $c^{-2}$, a systematic shift is observed between these two estimates, with $\chi_{\mathrm{micro}} \simeq 0.3 \, \chi_{\mathrm{macro}}$. This discrepancy may be attributed to several biases in the segmentation of tracks into runs. First, since only the two-dimensional projection of tracks can be measured, runs nearly normal to the measurement plane are perceived as low-velocity erratic motions and can be interpreted as tumbles. Second, the measurements are limited both in depth (the depth of field is \SI{30}{\um}) and acquisition time (image sequences are limited to \SI{5}{\s} to allow for a fast scan along the $x$-direction). Longer tracks are therefore truncated, resulting in a large number of small tracks without tumbles. This leads to an over-representation of short runs, and hence an underestimation of $\alpha$. The aerotactic response $\chi_{\mathrm{micro}}$ computed from the run time modulation therefore underestimates the true aerotactic response, but its overall scaling $c^{-2}$ remains unaffected by this bias.

\section{Conclusion}
\label{sec:ccl}

In this paper, we use an original setup allowing for simultaneous measurements of oxygen concentration and bacteria tracking to investigate the scaling of the aerotactic response. This setup allows us to quantitatively monitor bacteria during their migration to an oxygen source, reproducing a natural environmental situation in the laboratory. The aerotactic response $\chi$ computed both macroscopically, from the concentration profiles of bacteria, and microscopically, from the bias in run times, yields consistent results in the form $\chi \propto c^{-2}$. To our knowledge, this is the first direct measurement of $\chi(c)$.

The scaling $\chi \propto c^{-2}$ is consistent with the models based on the biochemistry of bacterial membrane receptors~\cite{brown1974temporal,lapidus1976model,rivero1989transport} given by Eq.~\eqref{eq:Chi_Kd_c2} for a dissociation constant $K_d$ much smaller than the typical oxygen concentration. It differs from the "log-sensing" scaling $\chi \propto c^{-1}$  reported in recent experiments using stationary oxygen gradients~\cite{kalinin2009,menolascina2017logarithmic} with other bacteria. 
This variability may originate from a specific aerotactic response of the different bacterial species, which exhibit different oxygen receptors or swimming patterns ("run-and-tumble", "run-reverse" or "run-reverse and flick"~\cite{taktikos2013motility, grognot_2021_propellers}).

The simultaneous mapping of oxygen concentration using a fluorescent O$_2$-sensor together with the bacteria tracking is a very promising approach. This setup could be used in principle with more complex chemical stimuli, when aerotaxis competes with other chemotaxis, or in complex environments mimicking natural ecosystems.

\acknowledgments

We thank N. Busset, A. Gargasson and G. Lextrait for experimental help and fruitful discussions. This work is supported by the French National Research Agency (ANR) through the "Laboratoire d'Excellence Physics Atom Light Mater" (LabEx PALM) as part of the "Investissements d'Avenir" program (ANR-10-LABX-0039), and by the CNRS through the Mission for Transversal and Interdisciplinary Initiatives (MITI), part of the 80 Prime program (RootBac project).

%%%%%%%%%%%%%%%%%%%%%%%%%%%%%%%%%%%%%%%%
\appendix

\section{Oxygen measurements}
\label{sec:SM_O2}

Oxygen concentration maps are obtained from fluorescence microscopy images of Ruthenium, an organo-metallic complex quenched by oxygen. Ruthenium complex [ruthenium-tris(4,7 diphenyl-1, 10-phenanthroline) dichloride, also named Ru(dpp)$_3$Cl$_2$ or Ru(dpp)] is encapsulated in phospholipidic micelles (referred to as Ru-micelles),to make it soluble in water and biocompatible~\cite{douarche2009coli,morse2016}. We did not observe any influence of Ru-micelles on the bacterial motility. 

\begin{figure}[b]
\includegraphics[trim = 10mm 62mm 18mm 70mm, clip, width=0.48\textwidth, angle=0]{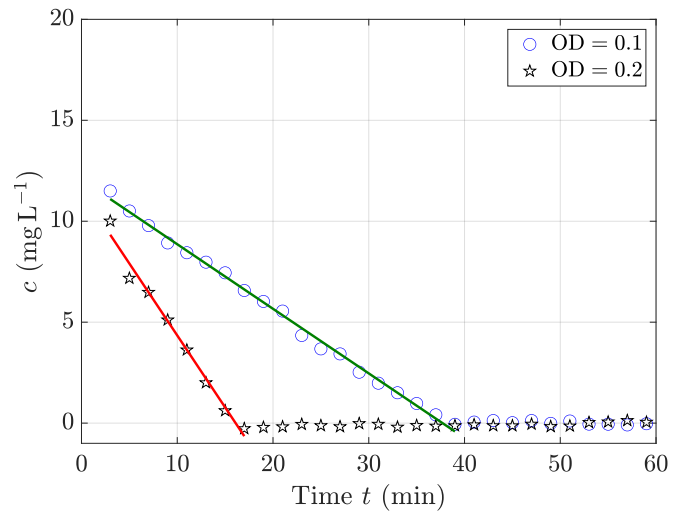} 
\caption{Temporal evolution of the dioxygen concentration $c$ in a suspension of \emph{B. contaminans} at $\mathrm{OD}=0.1$ (blue circles) and $\mathrm{OD}=0.2$ (black stars). The oxygen initially dissolved is consumed at a steady rate by the bacteria. The concentration is deduced from the fluorescence intensity using the Stern-Volmer relation [Eq.~\eqref{eq:Stern_volmer}]. The black line shows the linear fit $c(t)=c_0-k_c t$ with $c_0=\SI{11.8\pm3}{\mg\per\liter}$ and a consumption rate $k_c$ of 0.32 and \SI{0.71}{\mg\per\liter\per\minute}.}
\label{fig:SM_O2}
\end{figure}

Fluorescence intensity is converted into O$_2$ concentration using the Stern-Volmer relation
\begin{equation}
\label{eq:Stern_volmer}
    \dfrac{I_\mathrm{ref}}{I(c)} = 1 + K_q c,
\end{equation}
where $I(c)$ is the fluorescent intensity at oxygen concentration $c$, $I_\mathrm{ref}$ is the intensity without oxygen, and $K_q$ is the quenching constant. $I_\mathrm{ref}$ is taken as the highest intensity at the end of each experiment. This reference level was confirmed to be equal to $c=\SI{0}{\mg\per\liter}$ when measured with an electrochemical O$_2$-probe in a closed sample where bacteria consumed the available oxygen. Figure~\ref{fig:SM_O2} shows the temporal evolution of the oxygen concentration in two bacterial suspensions of different concentrations enclosed in a glass capillary with no oxygen supply. $K_q$ was determined by measuring the fluorescent intensity of a solution containing no oxygen (a solution saturated of \emph{B. contaminans} bacteria), and comparing it to the intensity level of a solution with a known oxygen concentration. We find $K_q=\SI{0.21}{\liter\per\mg}$, which is of the same order of magnitude of values reported for similar Ruthenium-based  dyes~\cite{gerritsen1997fluorescence,sud2006time,polinkovsky2009fine,kim2016oxygen,Adler2012,menolascina2017logarithmic}.

\section{"Run-reverse" swimming of \emph{B. contaminans}}
\label{sec:SM_run-reverse}

\begin{figure}[b]
\includegraphics[trim = 5mm 72mm 18mm 79mm, clip, width=0.48\textwidth, angle=0]{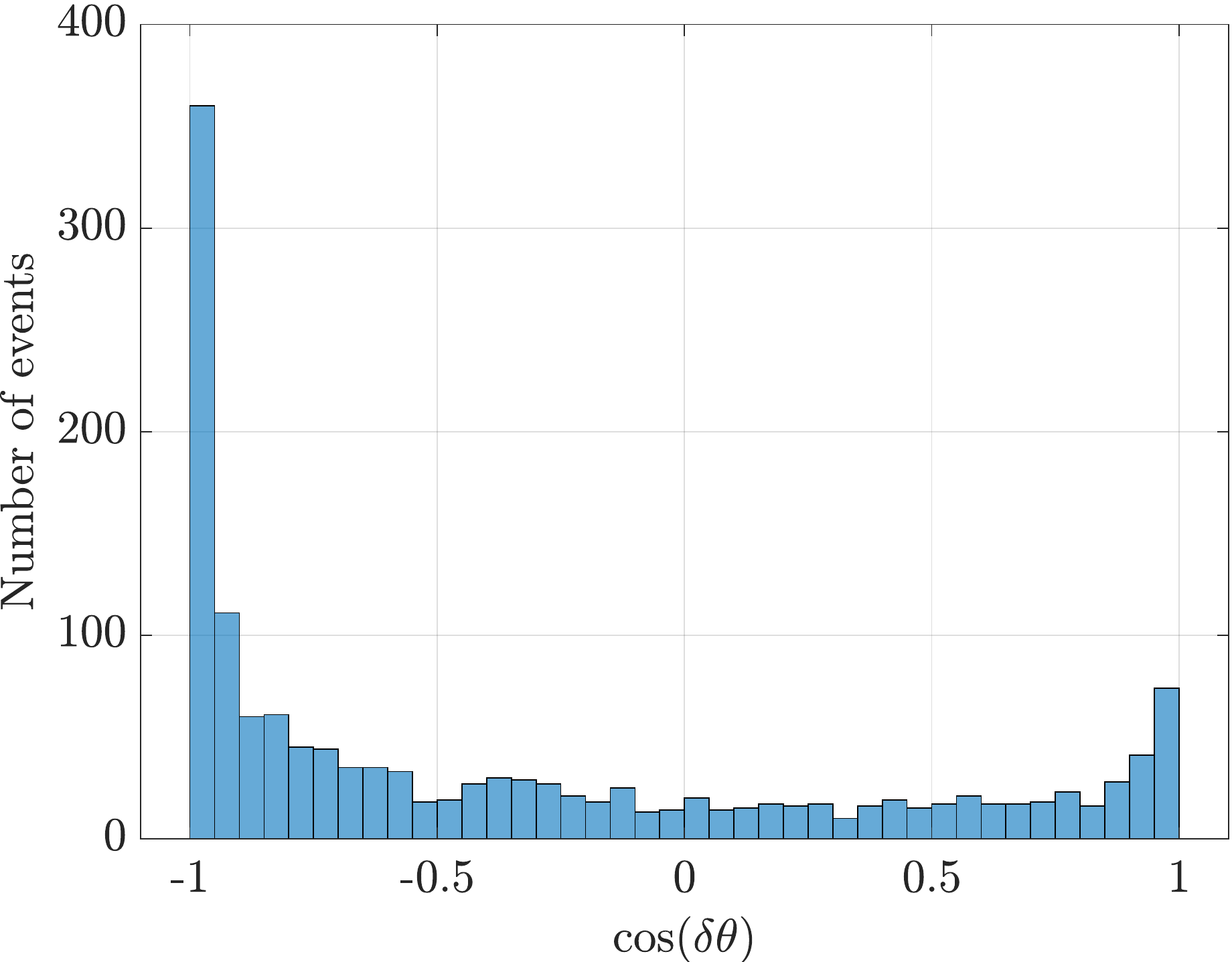} 
\caption{Distribution of the turning angle $\delta\theta$ between two successive runs. The distribution is peaked at $\cos\delta\theta=-1$, corresponding to reverses.}
\label{fig:SM_turning_angle_distri}
\end{figure}

\emph{B. contaminans} is a motile bacterium exhibiting a large dispersion of swimming velocities. Its velocity (projected on the microscope plane) ranges from 0 up to \SI{60}{\um\per\s}, with a mean value around \SIrange{20}{25}{\um\per\s}. A particular feature of \emph{B. contaminans} is its "run-reverse" swimming (illustrated in Fig.~\ref{fig:tracks}), similar to some marine bacteria~\cite{barbara2003bacterial,taktikos2013motility}. To quantify this run-reverse motion, we measure the turning angle $\delta\theta$ between two successive runs $j$ and $j+1$, defined as $\delta \theta = \theta_{\mathrm{run}, j+1} - \theta_{\mathrm{run},j}$. Its distribution, plotted in Fig.~\ref{fig:SM_turning_angle_distri}, shows a strong accumulation at $\cos \delta \theta \simeq -1$, i.e., a majority of tumbles correspond to reverses. We obtain a mean value of $\langle\cos\delta\theta\rangle\simeq -0.34$, much smaller than the value 0.33 found for the classical run-and-tumble motion in \emph{E. coli}~\cite{berg1993random}.

\section{Diffusion coefficient}
\label{sec:SM_mu}

\begin{figure}[htb]
\includegraphics[trim = 0mm 57mm 13mm 64mm, clip, width=0.48\textwidth, angle=0]{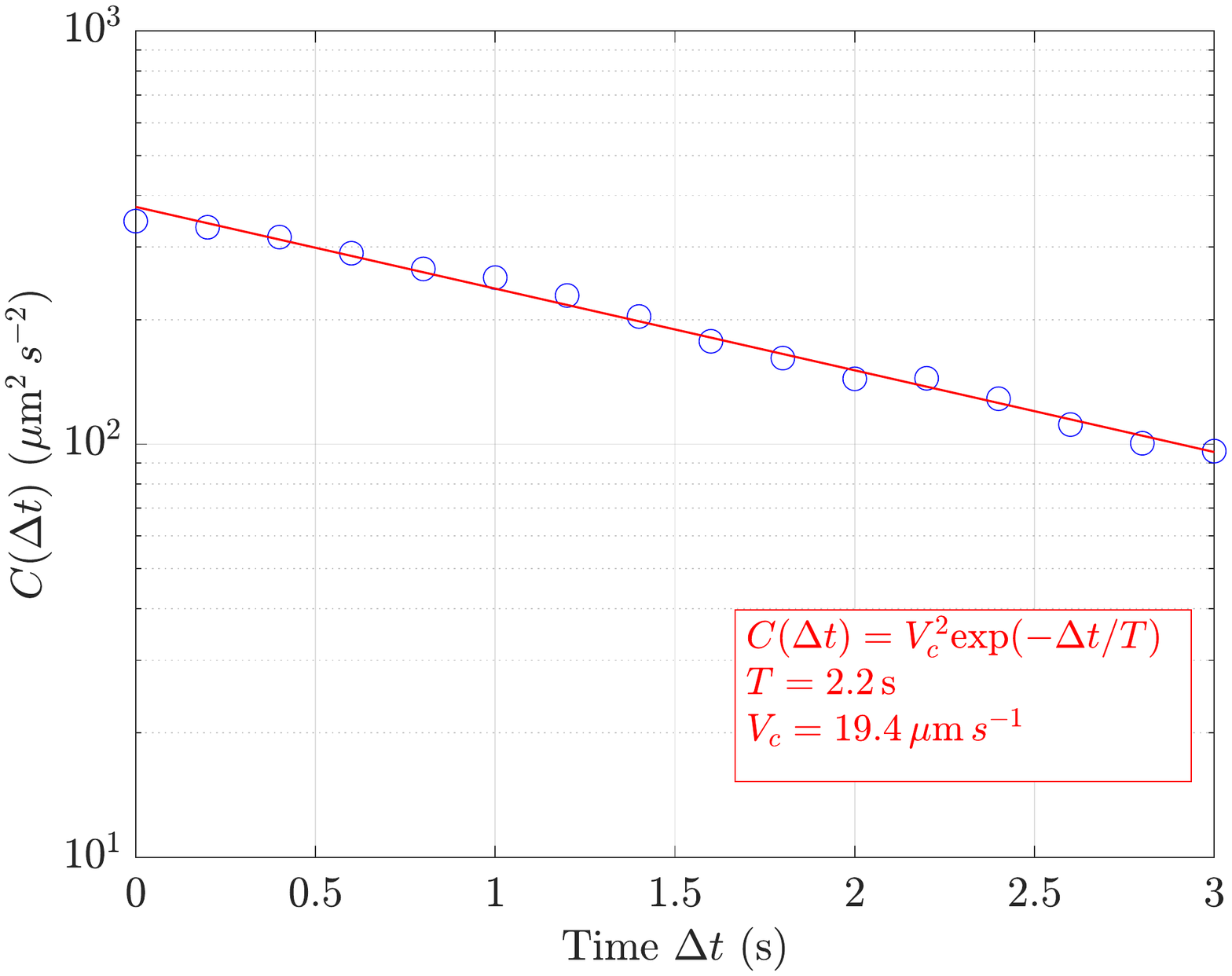} 
\caption{Velocity correlation function $C(\Delta t) = \langle {\bf V}(t) \cdot {\bf V} (t+\Delta t) \rangle$ at $\mathrm{OD}=0.05$. The red line shows the exponential fit [Eq.~\eqref{eq:fitC}], with $V_c=\SI{19.4}{\um\per\s}$ and $T=\SI{2.2}{\s}$.}
\label{fig:mu_correlation}
\end{figure}

To measure the diffusion coefficient $\mu$, we compute the velocity correlation function~\cite{lovely1975statistical,taktikos2013motility,lauga2020}
\begin{equation}
    C(\Delta t) = \langle {\bf V}(t) \cdot {\bf V} (t+\Delta t) \rangle
\end{equation}
averaged over all the bacterial tracks.  An example of $C(\Delta t)$ is shown in Fig.~\ref{fig:mu_correlation}. We fit the exponential decrease as  \begin{equation}
 C(\Delta t) \simeq V_c^2 \exp(-\Delta t/T),
 \label{eq:fitC}
\end{equation}
with $V_c = \sqrt{2/3} V_0$, and $V_0$ the mean swimming velocity,
from which we compute the diffusion coefficient as $\mu=(1/2) V_c^2 T$. An average over numerous experiments yields $\mu \simeq \SI{450\pm100}{\square\um\per\s}$.

\section{Aerotactic run time modulation}
\label{app:artm}

We examine here the aerotactic bias in the run time distribution. We first consider nonaerotactic run-and-tumble (in homogeneous oxygen concentration). The probability $p(\tau_{\mathrm{run}})$ that a bacteria makes a run of duration $\tau_{\mathrm{run}}$ is
\begin{equation}
    p(\tau_{\mathrm{run}})=p(\tau_{\mathrm{run}}-\delta t) p(\delta t).
    \label{eq:p}
\end{equation}
%\com{Here, $\tau_\mathrm{run}$ is a random variable which differs semantically from $\tau_\mathrm{run}(\theta_\mathrm{run})$ in the main text}.
Noting $\beta$ the probability per unit time of a tumble, we have $p(\delta t)= 1 - \beta \delta t$, and Eq.~\eqref{eq:p} becomes
\begin{equation}
\frac{p(\tau_{\mathrm{run}})-p(\tau_{\mathrm{run}}-\delta t)}{p(\tau_{\mathrm{run}} - \delta t)}=-\beta \delta t.
\end{equation}
The probability $p(\tau_{\mathrm{run}})$ is then given by
\begin{equation}
    \text{ln}(p(\tau_{\mathrm{run}}))=-\int_0^{\tau_{\mathrm{run}}} \beta dt + C,
    \label{eq:p(tau)}
\end{equation}
where $C$ is an integration constant defined such that  $\int_0^{+\infty}p(\tau_{\mathrm{run}})d\tau_{\mathrm{run}}=1$.
For constant $\beta$, the solution of Eq.~\eqref{eq:p(tau)} is a Poisson law, $p(\tau_{\mathrm{run}})=\beta \exp(-\beta \tau_{\mathrm{run}})$, and the average duration of a run is
\begin{equation}
\overline{\tau}_{\mathrm{run}}= \int_0^{\infty} \tau_{\mathrm{run}} p(\tau_{\mathrm{run}}) d\tau_{\mathrm{run}} = 1/\beta.
\label{Eq:tau}
\end{equation}

We consider now the run time distribution in the presence of an oxygen gradient. We derive 
the average of the runs of bacteria moving up ($\overline{\tau}^+_{\mathrm{run}}$) or down ($\overline{\tau}^-_{\mathrm{run}}$) the oxygen gradient under the assumptions that (i) the frequency $\beta$ depends only on the local oxygen concentration $c$, and (ii) the bacteria swim at a constant velocity $V_0$.

The probability to make a run of duration $\tau_{\mathrm{run}}$ is still given by Eq.~\eqref{eq:p(tau)} but with a frequency $\beta=\beta[c(t)]$ that depends on the local oxygen concentration.
The average duration of the runs of the bacteria moving up or down the gradient are
\begin{equation}
    \overline{\tau}^\pm_{\mathrm{run}}=
    \int_0^{\infty} \tau^\pm_{\mathrm{run}} p(\tau^\pm_{\mathrm{run}}) d\tau^\pm_{\mathrm{run}}.
    \label{eq:tau_moyen}
\end{equation}

A Taylor expansion of $\beta[c(t)]$ around its initial value at $t=0$ gives
\begin{equation}
    \beta[c(t)]=\beta[c(0)] + \frac{d\beta[c(t)]}{dt}\bigg|_{t=0} t,
    \end{equation}
yielding
\begin{equation}
\begin{split}
\text{ln}(p(\tau^{\pm}_\mathrm{run}))\simeq
-\int_0^{\tau^{\pm}_\mathrm{run}} \left( \beta[c(0)]+\frac{d\beta}{dt}\bigg|_{t=0} t \right) dt + C.%\\
%\simeq-(\beta(0)\tau + \frac{1}{2}\frac{d\beta}{dt}\bigg|_{t=0}\tau^2)+C\\
%\simeq- \tau^* (1+\frac{1}{2 \beta(0)^2}\frac{d \beta}{dt}\bigg|_{t=0} \tau^*) +C 
\end{split}
\end{equation}
For a constant swimming velocity in the $x$ direction, the time derivative of $\beta$ writes
$$\frac{d\beta[c(0)]}{dt}=V_0 \frac{d\beta[c(x)]}{dx},$$
where $x$ is the bacteria position at time $t=0$. We define the run-time modulation coefficient as
$$
\alpha(x) = \left| {2}\frac{V_0}{\beta[c(x)]^2}\frac{d\beta}{dx} \right|.
$$
In the limit of small $\alpha$, Eq.(\ref{eq:tau_moyen}) reduces to
\begin{equation}
    \overline{\tau}^{\pm}_{\mathrm{run}}=\overline{\tau}_{\mathrm{run}}
    \left(1\pm \alpha \right),
    \label{eq:average_tau}
\end{equation}
where $\overline{\tau}_{\mathrm{run}} =1/\beta(c)$ is the average run duration of bacteria moving in an homogeneous oxygen concentration $c$ [Eq.~\eqref{Eq:tau}]. We note that Eq.~\eqref{eq:average_tau} can be written in a way similar to Rivero \emph{et al.}~\cite{rivero1989transport} and Berg and Brown~\cite{berg1972chemotaxis},
\begin{equation}
    \text{ln}(\overline{\tau}^\pm_{\mathrm{run}}(a))=\text{ln}(\overline{\tau}_{\mathrm{run}})\pm \alpha.
    \label{eq:lntau}
\end{equation}

The calculation can be generalised by considering runs not aligned with the gradient, but making an angle $\theta_{\mathrm{run}}$ with respect to the gradient, yielding
\begin{equation}
    \overline{\tau}_{\mathrm{run}}(\theta_{\mathrm{run}})=\tau_{0} (1- \alpha \cos(\theta_{\mathrm{run}})),
    \label{eq:cos}
\end{equation}
where $\tau_{0}$ is the average run time measured normal to the gradient (for $\cos \theta_{\mathrm{run}}=0$). Here, we have $\theta_{\mathrm{run}}=\pi$ for runs moving up the gradient. 

The aerotactic velocity can be written 
%\begin{equation}
%v_a = -\frac{V_0}{\beta(c)} \left( \frac{1}{\overline{\tau}_{\mathrm{run}}^-}-\frac{1}{\overline{\tau}_{\mathrm{run}}^+} \right).
%\end{equation}
\begin{equation}
\label{eq:va_SM}
    v_a = \dfrac{1}{2\pi} \dfrac{1}{\tau_0} \int_0^{2\pi}\mathbf{v}\cdot \mathbf{e}_x \overline{\tau}_\mathrm{run}(\theta_\mathrm{run})d\theta_\mathrm{run}.
\end{equation}
Assuming $\tau_0$ and $V_0$ do not depend on the run angle, we finally obtain
\begin{equation}
    v_a = \dfrac{-\alpha V_0}{2}
\end{equation}
by combining Eqs.~\eqref{eq:cos} and \eqref{eq:va_SM}.

\bibliography{biblio_aerotaxis}% Produces the bibliography via BibTeX.

\begin{thebibliography}{66}%
\makeatletter
\providecommand \@ifxundefined [1]{%
 \@ifx{#1\undefined}
}%
\providecommand \@ifnum [1]{%
 \ifnum #1\expandafter \@firstoftwo
 \else \expandafter \@secondoftwo
 \fi
}%
\providecommand \@ifx [1]{%
 \ifx #1\expandafter \@firstoftwo
 \else \expandafter \@secondoftwo
 \fi
}%
\providecommand \natexlab [1]{#1}%
\providecommand \enquote  [1]{``#1''}%
\providecommand \bibnamefont  [1]{#1}%
\providecommand \bibfnamefont [1]{#1}%
\providecommand \citenamefont [1]{#1}%
\providecommand \href@noop [0]{\@secondoftwo}%
\providecommand \href [0]{\begingroup \@sanitize@url \@href}%
\providecommand \@href[1]{\@@startlink{#1}\@@href}%
\providecommand \@@href[1]{\endgroup#1\@@endlink}%
\providecommand \@sanitize@url [0]{\catcode `\\12\catcode `\$12\catcode
  `\&12\catcode `\#12\catcode `\^12\catcode `\_12\catcode `\%12\relax}%
\providecommand \@@startlink[1]{}%
\providecommand \@@endlink[0]{}%
\providecommand \url  [0]{\begingroup\@sanitize@url \@url }%
\providecommand \@url [1]{\endgroup\@href {#1}{\urlprefix }}%
\providecommand \urlprefix  [0]{URL }%
\providecommand \Eprint [0]{\href }%
\providecommand \doibase [0]{https://doi.org/}%
\providecommand \selectlanguage [0]{\@gobble}%
\providecommand \bibinfo  [0]{\@secondoftwo}%
\providecommand \bibfield  [0]{\@secondoftwo}%
\providecommand \translation [1]{[#1]}%
\providecommand \BibitemOpen [0]{}%
\providecommand \bibitemStop [0]{}%
\providecommand \bibitemNoStop [0]{.\EOS\space}%
\providecommand \EOS [0]{\spacefactor3000\relax}%
\providecommand \BibitemShut  [1]{\csname bibitem#1\endcsname}%
\let\auto@bib@innerbib\@empty
%</preamble>
\bibitem [{\citenamefont {Engelmann}(1881)}]{engelmann1881}%
  \BibitemOpen
  \bibfield  {author} {\bibinfo {author} {\bibfnamefont {T.~W.}\ \bibnamefont
  {Engelmann}},\ }\bibfield  {title} {\bibinfo {title} {Neue method zur
  untersuchung der sauerstoffausscheidung pflanzilcher und thierischer
  organismen},\ }\href@noop {} {\bibfield  {journal} {\bibinfo  {journal}
  {Botanische Zeitung}\ }\textbf {\bibinfo {volume} {39}},\ \bibinfo {pages}
  {441} (\bibinfo {year} {1881})}\BibitemShut {NoStop}%
\bibitem [{\citenamefont {Pfeffer}(1888)}]{pfeffer1887}%
  \BibitemOpen
  \bibfield  {author} {\bibinfo {author} {\bibfnamefont {W.}~\bibnamefont
  {Pfeffer}},\ }\bibfield  {title} {\bibinfo {title} {Über chemotaktische
  bewegungen von bakterien, flagellaten und volvocineen},\ }\href@noop {}
  {\bibfield  {journal} {\bibinfo  {journal} {Untersuch. bot. Inst. Tübingen}\
  }\textbf {\bibinfo {volume} {2}},\ \bibinfo {pages} {582} (\bibinfo {year}
  {1888})}\BibitemShut {NoStop}%
\bibitem [{\citenamefont {Kali-Cohen}(1890)}]{alicohen1890}%
  \BibitemOpen
  \bibfield  {author} {\bibinfo {author} {\bibfnamefont {C.~H.}\ \bibnamefont
  {Kali-Cohen}},\ }\bibfield  {title} {\bibinfo {title} {Die chemotaxis als
  hülfsmittelder bakteriologischen forschung},\ }\href@noop {} {\bibfield
  {journal} {\bibinfo  {journal} {Zentralbl. Bakteriol. Parasitenkd.
  Infektionskr Hyg. Abt.}\ }\textbf {\bibinfo {volume} {1}},\ \bibinfo {pages}
  {161} (\bibinfo {year} {1890})}\BibitemShut {NoStop}%
\bibitem [{\citenamefont {Beyerinck}(1893)}]{beyerinck1893}%
  \BibitemOpen
  \bibfield  {author} {\bibinfo {author} {\bibfnamefont {M.~W.}\ \bibnamefont
  {Beyerinck}},\ }\bibfield  {title} {\bibinfo {title} {Ueber atmungsfiguren
  beweglicher bakterien},\ }\href@noop {} {\bibfield  {journal} {\bibinfo
  {journal} {Zentrabl Bakteriol Parasitenkd}\ }\textbf {\bibinfo {volume}
  {14}},\ \bibinfo {pages} {827} (\bibinfo {year} {1893})}\BibitemShut
  {NoStop}%
\bibitem [{\citenamefont {Jennings}\ and\ \citenamefont
  {Crosby}(1901)}]{jennings1901studies}%
  \BibitemOpen
  \bibfield  {author} {\bibinfo {author} {\bibfnamefont {H.}~\bibnamefont
  {Jennings}}\ and\ \bibinfo {author} {\bibfnamefont {J.}~\bibnamefont
  {Crosby}},\ }\bibfield  {title} {\bibinfo {title} {Studies on reactions to
  stimuli in unicellular organisms.—vii. the manner in which bacteria react
  to stimuli, especially to chemical stimuli},\ }\href@noop {} {\bibfield
  {journal} {\bibinfo  {journal} {American Journal of Physiology-Legacy
  Content}\ }\textbf {\bibinfo {volume} {6}},\ \bibinfo {pages} {31} (\bibinfo
  {year} {1901})}\BibitemShut {NoStop}%
\bibitem [{\citenamefont {Adler}(1966{\natexlab{a}})}]{adler1966chemotaxis}%
  \BibitemOpen
  \bibfield  {author} {\bibinfo {author} {\bibfnamefont {J.}~\bibnamefont
  {Adler}},\ }\bibfield  {title} {\bibinfo {title} {Chemotaxis in bacteria},\
  }\href@noop {} {\bibfield  {journal} {\bibinfo  {journal} {Science}\ }\textbf
  {\bibinfo {volume} {153}},\ \bibinfo {pages} {708} (\bibinfo {year}
  {1966}{\natexlab{a}})}\BibitemShut {NoStop}%
\bibitem [{\citenamefont {Adler}(1966{\natexlab{b}})}]{adler1966effect}%
  \BibitemOpen
  \bibfield  {author} {\bibinfo {author} {\bibfnamefont {J.}~\bibnamefont
  {Adler}},\ }\bibfield  {title} {\bibinfo {title} {Effect of amino acids and
  oxygen on chemotaxis in \emph{Escherichia coli}},\ }\href@noop {} {\bibfield
  {journal} {\bibinfo  {journal} {Journal of bacteriology}\ }\textbf {\bibinfo
  {volume} {92}},\ \bibinfo {pages} {121} (\bibinfo {year}
  {1966}{\natexlab{b}})}\BibitemShut {NoStop}%
\bibitem [{\citenamefont {Taylor}\ \emph {et~al.}(1999)\citenamefont {Taylor},
  \citenamefont {Zhulin},\ and\ \citenamefont {Johnson}}]{Taylor1999}%
  \BibitemOpen
  \bibfield  {author} {\bibinfo {author} {\bibfnamefont {B.~L.}\ \bibnamefont
  {Taylor}}, \bibinfo {author} {\bibfnamefont {I.~B.}\ \bibnamefont {Zhulin}},\
  and\ \bibinfo {author} {\bibfnamefont {M.~S.}\ \bibnamefont {Johnson}},\
  }\bibfield  {title} {\bibinfo {title} {Aerotaxis and other energy-sensing
  behavior in bacteria},\ }\href
  {https://doi.org/10.1146/annurev.micro.53.1.103} {\bibfield  {journal}
  {\bibinfo  {journal} {Annual Review of Microbiology}\ }\textbf {\bibinfo
  {volume} {53}},\ \bibinfo {pages} {103} (\bibinfo {year} {1999})},\ \bibinfo
  {note} {pMID: 10547687}\BibitemShut {NoStop}%
\bibitem [{\citenamefont {Berg}\ and\ \citenamefont
  {Brown}(1972)}]{berg1972chemotaxis}%
  \BibitemOpen
  \bibfield  {author} {\bibinfo {author} {\bibfnamefont {H.~C.}\ \bibnamefont
  {Berg}}\ and\ \bibinfo {author} {\bibfnamefont {D.~A.}\ \bibnamefont
  {Brown}},\ }\bibfield  {title} {\bibinfo {title} {Chemotaxis in
  \emph{Escherichia coli} analysed by three-dimensional tracking},\ }\href@noop
  {} {\bibfield  {journal} {\bibinfo  {journal} {Nature}\ }\textbf {\bibinfo
  {volume} {239}},\ \bibinfo {pages} {500} (\bibinfo {year}
  {1972})}\BibitemShut {NoStop}%
\bibitem [{\citenamefont {Brown}\ and\ \citenamefont
  {Berg}(1974)}]{brown1974temporal}%
  \BibitemOpen
  \bibfield  {author} {\bibinfo {author} {\bibfnamefont {D.~A.}\ \bibnamefont
  {Brown}}\ and\ \bibinfo {author} {\bibfnamefont {H.~C.}\ \bibnamefont
  {Berg}},\ }\bibfield  {title} {\bibinfo {title} {Temporal stimulation of
  chemotaxis in \emph{Escherichia coli}},\ }\href@noop {} {\bibfield  {journal}
  {\bibinfo  {journal} {Proceedings of the National Academy of Sciences}\
  }\textbf {\bibinfo {volume} {71}},\ \bibinfo {pages} {1388} (\bibinfo {year}
  {1974})}\BibitemShut {NoStop}%
\bibitem [{\citenamefont {Dahlquist}\ \emph {et~al.}(1972)\citenamefont
  {Dahlquist}, \citenamefont {Lovely},\ and\ \citenamefont
  {Koshland}}]{dahlquist1972quantitative}%
  \BibitemOpen
  \bibfield  {author} {\bibinfo {author} {\bibfnamefont {F.}~\bibnamefont
  {Dahlquist}}, \bibinfo {author} {\bibfnamefont {P.}~\bibnamefont {Lovely}},\
  and\ \bibinfo {author} {\bibfnamefont {D.}~\bibnamefont {Koshland}},\
  }\bibfield  {title} {\bibinfo {title} {Quantitative analysis of bacterial
  migration in chemotaxis},\ }\href@noop {} {\bibfield  {journal} {\bibinfo
  {journal} {Nature New Biology}\ }\textbf {\bibinfo {volume} {236}},\ \bibinfo
  {pages} {120} (\bibinfo {year} {1972})}\BibitemShut {NoStop}%
\bibitem [{\citenamefont {Ford}\ and\ \citenamefont
  {Lauffenburger}(1991)}]{ford1991measurement2}%
  \BibitemOpen
  \bibfield  {author} {\bibinfo {author} {\bibfnamefont {R.~M.}\ \bibnamefont
  {Ford}}\ and\ \bibinfo {author} {\bibfnamefont {D.~A.}\ \bibnamefont
  {Lauffenburger}},\ }\bibfield  {title} {\bibinfo {title} {Measurement of
  bacterial random motility and chemotaxis coefficients: Ii. application of
  single-cell-based mathematical model},\ }\href@noop {} {\bibfield  {journal}
  {\bibinfo  {journal} {Biotechnology and bioengineering}\ }\textbf {\bibinfo
  {volume} {37}},\ \bibinfo {pages} {661} (\bibinfo {year} {1991})}\BibitemShut
  {NoStop}%
\bibitem [{\citenamefont {Lewus}\ and\ \citenamefont
  {Ford}(2001)}]{lewus2001quantification}%
  \BibitemOpen
  \bibfield  {author} {\bibinfo {author} {\bibfnamefont {P.}~\bibnamefont
  {Lewus}}\ and\ \bibinfo {author} {\bibfnamefont {R.~M.}\ \bibnamefont
  {Ford}},\ }\bibfield  {title} {\bibinfo {title} {Quantification of random
  motility and chemotaxis bacterial transport coefficients using
  individual-cell and population-scale assays},\ }\href@noop {} {\bibfield
  {journal} {\bibinfo  {journal} {Biotechnology and bioengineering}\ }\textbf
  {\bibinfo {volume} {75}},\ \bibinfo {pages} {292} (\bibinfo {year}
  {2001})}\BibitemShut {NoStop}%
\bibitem [{\citenamefont {G{\"o}tz}\ \emph {et~al.}(1982)\citenamefont
  {G{\"o}tz}, \citenamefont {Limmer}, \citenamefont {Ober},\ and\ \citenamefont
  {Schmitt}}]{gotz1982motility}%
  \BibitemOpen
  \bibfield  {author} {\bibinfo {author} {\bibfnamefont {R.}~\bibnamefont
  {G{\"o}tz}}, \bibinfo {author} {\bibfnamefont {N.}~\bibnamefont {Limmer}},
  \bibinfo {author} {\bibfnamefont {K.}~\bibnamefont {Ober}},\ and\ \bibinfo
  {author} {\bibfnamefont {R.}~\bibnamefont {Schmitt}},\ }\bibfield  {title}
  {\bibinfo {title} {Motility and chemotaxis in two strains of \emph{Rhizobium}
  with complex flagella},\ }\href@noop {} {\bibfield  {journal} {\bibinfo
  {journal} {Microbiology}\ }\textbf {\bibinfo {volume} {128}},\ \bibinfo
  {pages} {789} (\bibinfo {year} {1982})}\BibitemShut {NoStop}%
\bibitem [{\citenamefont {G{\"o}tz}\ and\ \citenamefont
  {Schmitt}(1987)}]{gotz1987rhizobium}%
  \BibitemOpen
  \bibfield  {author} {\bibinfo {author} {\bibfnamefont {R.}~\bibnamefont
  {G{\"o}tz}}\ and\ \bibinfo {author} {\bibfnamefont {R.}~\bibnamefont
  {Schmitt}},\ }\bibfield  {title} {\bibinfo {title} {\emph{Rhizobium meliloti}
  swims by unidirectional, intermittent rotation of right-handed flagellar
  helices.},\ }\href@noop {} {\bibfield  {journal} {\bibinfo  {journal}
  {Journal of bacteriology}\ }\textbf {\bibinfo {volume} {169}},\ \bibinfo
  {pages} {3146} (\bibinfo {year} {1987})}\BibitemShut {NoStop}%
\bibitem [{\citenamefont {Baracchini}\ and\ \citenamefont
  {Sherris}(1959)}]{Baracchini1959}%
  \BibitemOpen
  \bibfield  {author} {\bibinfo {author} {\bibfnamefont {O.}~\bibnamefont
  {Baracchini}}\ and\ \bibinfo {author} {\bibfnamefont {J.~C.}\ \bibnamefont
  {Sherris}},\ }\bibfield  {title} {\bibinfo {title} {The chemotactic effect of
  oxygen on bacteria},\ }\href
  {https://doi.org/https://doi.org/10.1002/path.1700770228} {\bibfield
  {journal} {\bibinfo  {journal} {The Journal of Pathology and Bacteriology}\
  }\textbf {\bibinfo {volume} {77}},\ \bibinfo {pages} {565} (\bibinfo {year}
  {1959})}\BibitemShut {NoStop}%
\bibitem [{\citenamefont {Mazzag}\ \emph {et~al.}(2003)\citenamefont {Mazzag},
  \citenamefont {Zhulin},\ and\ \citenamefont {Mogilner}}]{mazzag2003model}%
  \BibitemOpen
  \bibfield  {author} {\bibinfo {author} {\bibfnamefont {B.}~\bibnamefont
  {Mazzag}}, \bibinfo {author} {\bibfnamefont {I.}~\bibnamefont {Zhulin}},\
  and\ \bibinfo {author} {\bibfnamefont {A.}~\bibnamefont {Mogilner}},\
  }\bibfield  {title} {\bibinfo {title} {Model of bacterial band formation in
  aerotaxis},\ }\href@noop {} {\bibfield  {journal} {\bibinfo  {journal}
  {Biophysical journal}\ }\textbf {\bibinfo {volume} {85}},\ \bibinfo {pages}
  {3558} (\bibinfo {year} {2003})}\BibitemShut {NoStop}%
\bibitem [{\citenamefont {Stricker}\ \emph {et~al.}(2020)\citenamefont
  {Stricker}, \citenamefont {Guido}, \citenamefont {Breithaupt}, \citenamefont
  {Mazza},\ and\ \citenamefont {Vollmer}}]{Stricker2020}%
  \BibitemOpen
  \bibfield  {author} {\bibinfo {author} {\bibfnamefont {L.}~\bibnamefont
  {Stricker}}, \bibinfo {author} {\bibfnamefont {I.}~\bibnamefont {Guido}},
  \bibinfo {author} {\bibfnamefont {T.}~\bibnamefont {Breithaupt}}, \bibinfo
  {author} {\bibfnamefont {M.~G.}\ \bibnamefont {Mazza}},\ and\ \bibinfo
  {author} {\bibfnamefont {J.}~\bibnamefont {Vollmer}},\ }\bibfield  {title}
  {\bibinfo {title} {Hybrid sideways/longitudinal swimming in the
  monoflagellate \emph{Shewanella oneidensis}: from aerotactic band to
  biofilm},\ }\href {https://doi.org/10.1098/rsif.2020.0559} {\bibfield
  {journal} {\bibinfo  {journal} {Journal of The Royal Society Interface}\
  }\textbf {\bibinfo {volume} {17}},\ \bibinfo {pages} {20200559} (\bibinfo
  {year} {2020})}\BibitemShut {NoStop}%
\bibitem [{\citenamefont {Frankel}\ \emph {et~al.}(1997)\citenamefont
  {Frankel}, \citenamefont {Bazylinski}, \citenamefont {Johnson},\ and\
  \citenamefont {Taylor}}]{Frankel1997}%
  \BibitemOpen
  \bibfield  {author} {\bibinfo {author} {\bibfnamefont {R.~B.}\ \bibnamefont
  {Frankel}}, \bibinfo {author} {\bibfnamefont {D.~A.}\ \bibnamefont
  {Bazylinski}}, \bibinfo {author} {\bibfnamefont {M.~S.}\ \bibnamefont
  {Johnson}},\ and\ \bibinfo {author} {\bibfnamefont {B.~L.}\ \bibnamefont
  {Taylor}},\ }\bibfield  {title} {\bibinfo {title} {Magneto-aerotaxis in
  marine coccoid bacteria},\ }\href@noop {} {\bibfield  {journal} {\bibinfo
  {journal} {Biophysical Journal}\ }\textbf {\bibinfo {volume} {73}} (\bibinfo
  {year} {1997})}\BibitemShut {NoStop}%
\bibitem [{\citenamefont {Popp}\ \emph {et~al.}(2014)\citenamefont {Popp},
  \citenamefont {Armitage},\ and\ \citenamefont
  {Sch{\"u}ler}}]{popp2014polarity}%
  \BibitemOpen
  \bibfield  {author} {\bibinfo {author} {\bibfnamefont {F.}~\bibnamefont
  {Popp}}, \bibinfo {author} {\bibfnamefont {J.~P.}\ \bibnamefont {Armitage}},\
  and\ \bibinfo {author} {\bibfnamefont {D.}~\bibnamefont {Sch{\"u}ler}},\
  }\bibfield  {title} {\bibinfo {title} {Polarity of bacterial magnetotaxis is
  controlled by aerotaxis through a common sensory pathway},\ }\href@noop {}
  {\bibfield  {journal} {\bibinfo  {journal} {Nature communications}\ }\textbf
  {\bibinfo {volume} {5}},\ \bibinfo {pages} {1} (\bibinfo {year}
  {2014})}\BibitemShut {NoStop}%
\bibitem [{\citenamefont {Fischer}\ and\ \citenamefont
  {Cypionka}(2006)}]{Fischer2006}%
  \BibitemOpen
  \bibfield  {author} {\bibinfo {author} {\bibfnamefont {J.~P.}\ \bibnamefont
  {Fischer}}\ and\ \bibinfo {author} {\bibfnamefont {H.}~\bibnamefont
  {Cypionka}},\ }\bibfield  {title} {\bibinfo {title} {{Analysis of aerotactic
  band formation by \emph{Desulfovibrio desulfuricans} in a stopped-flow
  diffusion chamber}},\ }\href
  {https://doi.org/10.1111/j.1574-695X.2005.00024.x} {\bibfield  {journal}
  {\bibinfo  {journal} {FEMS Microbiology Ecology}\ }\textbf {\bibinfo {volume}
  {55}},\ \bibinfo {pages} {186} (\bibinfo {year} {2006})}\BibitemShut
  {NoStop}%
\bibitem [{\citenamefont {Menolascina}\ \emph {et~al.}(2017)\citenamefont
  {Menolascina}, \citenamefont {Rusconi}, \citenamefont {Fernandez},
  \citenamefont {Smriga}, \citenamefont {Aminzare}, \citenamefont {Sontag},\
  and\ \citenamefont {Stocker}}]{menolascina2017logarithmic}%
  \BibitemOpen
  \bibfield  {author} {\bibinfo {author} {\bibfnamefont {F.}~\bibnamefont
  {Menolascina}}, \bibinfo {author} {\bibfnamefont {R.}~\bibnamefont
  {Rusconi}}, \bibinfo {author} {\bibfnamefont {V.~I.}\ \bibnamefont
  {Fernandez}}, \bibinfo {author} {\bibfnamefont {S.}~\bibnamefont {Smriga}},
  \bibinfo {author} {\bibfnamefont {Z.}~\bibnamefont {Aminzare}}, \bibinfo
  {author} {\bibfnamefont {E.~D.}\ \bibnamefont {Sontag}},\ and\ \bibinfo
  {author} {\bibfnamefont {R.}~\bibnamefont {Stocker}},\ }\bibfield  {title}
  {\bibinfo {title} {Logarithmic sensing in \emph{Bacillus subtilis}
  aerotaxis},\ }\href@noop {} {\bibfield  {journal} {\bibinfo  {journal} {NPJ
  systems biology and applications}\ }\textbf {\bibinfo {volume} {3}},\
  \bibinfo {pages} {1} (\bibinfo {year} {2017})}\BibitemShut {NoStop}%
\bibitem [{\citenamefont {Adler}\ \emph {et~al.}(2012)\citenamefont {Adler},
  \citenamefont {Erickstad}, \citenamefont {Gutierrez},\ and\ \citenamefont
  {Groisman}}]{Adler2012}%
  \BibitemOpen
  \bibfield  {author} {\bibinfo {author} {\bibfnamefont {M.}~\bibnamefont
  {Adler}}, \bibinfo {author} {\bibfnamefont {M.}~\bibnamefont {Erickstad}},
  \bibinfo {author} {\bibfnamefont {E.}~\bibnamefont {Gutierrez}},\ and\
  \bibinfo {author} {\bibfnamefont {A.}~\bibnamefont {Groisman}},\ }\bibfield
  {title} {\bibinfo {title} {Studies of bacterial aerotaxis in a microfluidic
  device},\ }\href {https://doi.org/10.1039/C2LC21006A} {\bibfield  {journal}
  {\bibinfo  {journal} {Lab Chip}\ }\textbf {\bibinfo {volume} {12}},\ \bibinfo
  {pages} {4835} (\bibinfo {year} {2012})}\BibitemShut {NoStop}%
\bibitem [{\citenamefont {Morse}\ \emph {et~al.}(2016)\citenamefont {Morse},
  \citenamefont {Colin}, \citenamefont {Wilson},\ and\ \citenamefont
  {Tang}}]{morse2016}%
  \BibitemOpen
  \bibfield  {author} {\bibinfo {author} {\bibfnamefont {M.}~\bibnamefont
  {Morse}}, \bibinfo {author} {\bibfnamefont {R.}~\bibnamefont {Colin}},
  \bibinfo {author} {\bibfnamefont {L.}~\bibnamefont {Wilson}},\ and\ \bibinfo
  {author} {\bibfnamefont {J.}~\bibnamefont {Tang}},\ }\bibfield  {title}
  {\bibinfo {title} {The aerotactic response of \emph{Caulobacter
  crescentus}},\ }\href {https://doi.org/10.1016/j.bpj.2016.03.028} {\bibfield
  {journal} {\bibinfo  {journal} {Biophysical Journal}\ }\textbf {\bibinfo
  {volume} {110}},\ \bibinfo {pages} {2076} (\bibinfo {year}
  {2016})}\BibitemShut {NoStop}%
\bibitem [{\citenamefont {Taylor}(1983)}]{Taylor1983}%
  \BibitemOpen
  \bibfield  {author} {\bibinfo {author} {\bibfnamefont {B.~L.}\ \bibnamefont
  {Taylor}},\ }\bibfield  {title} {\bibinfo {title} {How do bacteria find the
  optimal concentration of oxygen?},\ }\href
  {https://doi.org/10.1016/0968-0004(83)90030-0} {\bibfield  {journal}
  {\bibinfo  {journal} {Trends in Biochemical Sciences}\ }\textbf {\bibinfo
  {volume} {8}},\ \bibinfo {pages} {438} (\bibinfo {year} {1983})}\BibitemShut
  {NoStop}%
\bibitem [{\citenamefont {Armitano}\ \emph {et~al.}(2013)\citenamefont
  {Armitano}, \citenamefont {M\'ejean},\ and\ \citenamefont
  {Jourlin-Castelli}}]{Armitano2013}%
  \BibitemOpen
  \bibfield  {author} {\bibinfo {author} {\bibfnamefont {J.}~\bibnamefont
  {Armitano}}, \bibinfo {author} {\bibfnamefont {V.}~\bibnamefont {M\'ejean}},\
  and\ \bibinfo {author} {\bibfnamefont {C.}~\bibnamefont {Jourlin-Castelli}},\
  }\bibfield  {title} {\bibinfo {title} {Aerotaxis governs floating biofilm
  formation in \emph{Shewanella oneidensis}},\ }\href
  {https://doi.org/10.1111/1462-2920.12158} {\bibfield  {journal} {\bibinfo
  {journal} {Environmental Microbiology}\ }\textbf {\bibinfo {volume} {15}},\
  \bibinfo {pages} {3108} (\bibinfo {year} {2013})}\BibitemShut {NoStop}%
\bibitem [{\citenamefont {Ardr{\'{e}}}\ \emph {et~al.}(2015)\citenamefont
  {Ardr{\'{e}}}, \citenamefont {Henry}, \citenamefont {Douarche},\ and\
  \citenamefont {Plapp}}]{Ardre2015}%
  \BibitemOpen
  \bibfield  {author} {\bibinfo {author} {\bibfnamefont {M.}~\bibnamefont
  {Ardr{\'{e}}}}, \bibinfo {author} {\bibfnamefont {H.}~\bibnamefont {Henry}},
  \bibinfo {author} {\bibfnamefont {C.}~\bibnamefont {Douarche}},\ and\
  \bibinfo {author} {\bibfnamefont {M.}~\bibnamefont {Plapp}},\ }\bibfield
  {title} {\bibinfo {title} {An individual-based model for biofilm formation at
  liquid surfaces},\ }\href {https://doi.org/10.1088/1478-3975/12/6/066015}
  {\bibfield  {journal} {\bibinfo  {journal} {Physical Biology}\ }\textbf
  {\bibinfo {volume} {12}},\ \bibinfo {pages} {066015} (\bibinfo {year}
  {2015})}\BibitemShut {NoStop}%
\bibitem [{\citenamefont {Holscher}\ \emph {et~al.}(2015)\citenamefont
  {Holscher}, \citenamefont {Bartels}, \citenamefont {Lin}, \citenamefont
  {Gallegos-Monterrosa}, \citenamefont {Price-Whelan}, \citenamefont {Kolter},
  \citenamefont {Dietrich},\ and\ \citenamefont {Kovacs}}]{Holscher2015}%
  \BibitemOpen
  \bibfield  {author} {\bibinfo {author} {\bibfnamefont {T.}~\bibnamefont
  {Holscher}}, \bibinfo {author} {\bibfnamefont {B.}~\bibnamefont {Bartels}},
  \bibinfo {author} {\bibfnamefont {Y.-C.}\ \bibnamefont {Lin}}, \bibinfo
  {author} {\bibfnamefont {R.}~\bibnamefont {Gallegos-Monterrosa}}, \bibinfo
  {author} {\bibfnamefont {A.}~\bibnamefont {Price-Whelan}}, \bibinfo {author}
  {\bibfnamefont {R.}~\bibnamefont {Kolter}}, \bibinfo {author} {\bibfnamefont
  {L.~E.~P.}\ \bibnamefont {Dietrich}},\ and\ \bibinfo {author} {\bibfnamefont
  {A.~T.}\ \bibnamefont {Kovacs}},\ }\bibfield  {title} {\bibinfo {title}
  {Motility, chemotaxis and aerotaxis contribute to competitiveness during
  bacterial pellicle biofilm development.},\ }\href
  {https://doi.org/10.1016/j.jmb.2015.06.014.} {\bibfield  {journal} {\bibinfo
  {journal} {Journal of Molecular Biology}\ }\textbf {\bibinfo {volume}
  {427}},\ \bibinfo {pages} {3695} (\bibinfo {year} {2015})}\BibitemShut
  {NoStop}%
\bibitem [{\citenamefont {Wu}\ and\ \citenamefont {Jin}(2019)}]{Wu2019}%
  \BibitemOpen
  \bibfield  {author} {\bibinfo {author} {\bibfnamefont {G.}~\bibnamefont
  {Wu}}\ and\ \bibinfo {author} {\bibfnamefont {F.}~\bibnamefont {Jin}},\
  }\bibfield  {title} {\bibinfo {title} {Pellicle development of
  \emph{Shewanella oneidensis} is an aerotaxis-piloted and energy-dependent
  process},\ }\href
  {https://doi.org/https://doi.org/10.1016/j.bbrc.2019.08.144} {\bibfield
  {journal} {\bibinfo  {journal} {Biochemical and Biophysical Research
  Communications}\ }\textbf {\bibinfo {volume} {519}},\ \bibinfo {pages} {127 }
  (\bibinfo {year} {2019})}\BibitemShut {NoStop}%
\bibitem [{\citenamefont {Alexandre}\ \emph {et~al.}(2004)\citenamefont
  {Alexandre}, \citenamefont {Greer-Phillips},\ and\ \citenamefont
  {Zhulin}}]{alexandre_2004}%
  \BibitemOpen
  \bibfield  {author} {\bibinfo {author} {\bibfnamefont {G.}~\bibnamefont
  {Alexandre}}, \bibinfo {author} {\bibfnamefont {S.}~\bibnamefont
  {Greer-Phillips}},\ and\ \bibinfo {author} {\bibfnamefont {I.}~\bibnamefont
  {Zhulin}},\ }\bibfield  {title} {\bibinfo {title} {Ecological role of energy
  taxis in microorganisms},\ }\bibfield  {journal} {\bibinfo  {journal} {FEMS
  Microbiology Reviews}\ }\textbf {\bibinfo {volume} {28}},\ \href
  {https://doi.org/10.1016/j.femsre.2003.10.003} {10.1016/j.femsre.2003.10.003}
  (\bibinfo {year} {2004})\BibitemShut {NoStop}%
\bibitem [{\citenamefont {Tamar}\ \emph {et~al.}(2016)\citenamefont {Tamar},
  \citenamefont {Koler},\ and\ \citenamefont {Vaknin}}]{Tamar2016}%
  \BibitemOpen
  \bibfield  {author} {\bibinfo {author} {\bibfnamefont {E.}~\bibnamefont
  {Tamar}}, \bibinfo {author} {\bibfnamefont {M.}~\bibnamefont {Koler}},\ and\
  \bibinfo {author} {\bibfnamefont {A.}~\bibnamefont {Vaknin}},\ }\bibfield
  {title} {\bibinfo {title} {The role of motility and chemotaxis in the
  bacterial colonization of protected surfaces},\ }\bibfield  {journal}
  {\bibinfo  {journal} {Scientific Reports}\ }\textbf {\bibinfo {volume} {6}},\
  \href {https://doi.org/10.1038/srep19616} {10.1038/srep19616} (\bibinfo
  {year} {2016})\BibitemShut {NoStop}%
\bibitem [{\citenamefont {Felfoul}\ \emph {et~al.}(2016)\citenamefont
  {Felfoul}, \citenamefont {Mohammadi}, \citenamefont {Taherkhani},
  \citenamefont {de~Lanauze}, \citenamefont {Zhong~Xu}, \citenamefont {Loghin},
  \citenamefont {Essa}, \citenamefont {Jancik}, \citenamefont {Houle},
  \citenamefont {Lafleur}, \citenamefont {Gaboury}, \citenamefont {Tabrizian},
  \citenamefont {Kaou}, \citenamefont {Atkin}, \citenamefont {Vuong},
  \citenamefont {Batist}, \citenamefont {Beauchemin}, \citenamefont
  {Radzioch},\ and\ \citenamefont {Martel}}]{felfoul_magneto-aerotactic_2016}%
  \BibitemOpen
  \bibfield  {author} {\bibinfo {author} {\bibfnamefont {O.}~\bibnamefont
  {Felfoul}}, \bibinfo {author} {\bibfnamefont {M.}~\bibnamefont {Mohammadi}},
  \bibinfo {author} {\bibfnamefont {S.}~\bibnamefont {Taherkhani}}, \bibinfo
  {author} {\bibfnamefont {D.}~\bibnamefont {de~Lanauze}}, \bibinfo {author}
  {\bibfnamefont {Y.}~\bibnamefont {Zhong~Xu}}, \bibinfo {author}
  {\bibfnamefont {D.}~\bibnamefont {Loghin}}, \bibinfo {author} {\bibfnamefont
  {S.}~\bibnamefont {Essa}}, \bibinfo {author} {\bibfnamefont {S.}~\bibnamefont
  {Jancik}}, \bibinfo {author} {\bibfnamefont {D.}~\bibnamefont {Houle}},
  \bibinfo {author} {\bibfnamefont {M.}~\bibnamefont {Lafleur}}, \bibinfo
  {author} {\bibfnamefont {L.}~\bibnamefont {Gaboury}}, \bibinfo {author}
  {\bibfnamefont {M.}~\bibnamefont {Tabrizian}}, \bibinfo {author}
  {\bibfnamefont {N.}~\bibnamefont {Kaou}}, \bibinfo {author} {\bibfnamefont
  {M.}~\bibnamefont {Atkin}}, \bibinfo {author} {\bibfnamefont
  {T.}~\bibnamefont {Vuong}}, \bibinfo {author} {\bibfnamefont
  {G.}~\bibnamefont {Batist}}, \bibinfo {author} {\bibfnamefont
  {N.}~\bibnamefont {Beauchemin}}, \bibinfo {author} {\bibfnamefont
  {D.}~\bibnamefont {Radzioch}},\ and\ \bibinfo {author} {\bibfnamefont
  {S.}~\bibnamefont {Martel}},\ }\bibfield  {title} {\bibinfo {title}
  {Magneto-aerotactic bacteria deliver drug-containing nanoliposomes to tumour
  hypoxic regions},\ }\href {https://doi.org/10.1038/nnano.2016.137} {\bibfield
   {journal} {\bibinfo  {journal} {Nature Nanotechnology}\ }\textbf {\bibinfo
  {volume} {11}},\ \bibinfo {pages} {941} (\bibinfo {year} {2016})}\BibitemShut
  {NoStop}%
\bibitem [{\citenamefont {Baban}\ \emph {et~al.}(2010)\citenamefont {Baban},
  \citenamefont {Cronin}, \citenamefont {O’Hanlon}, \citenamefont
  {O’Sullivan},\ and\ \citenamefont {Tangney}}]{baban_bacteria_2010}%
  \BibitemOpen
  \bibfield  {author} {\bibinfo {author} {\bibfnamefont {C.~K.}\ \bibnamefont
  {Baban}}, \bibinfo {author} {\bibfnamefont {M.}~\bibnamefont {Cronin}},
  \bibinfo {author} {\bibfnamefont {D.}~\bibnamefont {O’Hanlon}}, \bibinfo
  {author} {\bibfnamefont {G.~C.}\ \bibnamefont {O’Sullivan}},\ and\ \bibinfo
  {author} {\bibfnamefont {M.}~\bibnamefont {Tangney}},\ }\bibfield  {title}
  {\bibinfo {title} {Bacteria as vectors for gene therapy of cancer},\ }\href
  {https://doi.org/10.4161/bbug.1.6.13146} {\bibfield  {journal} {\bibinfo
  {journal} {Bioengineered Bugs}\ }\textbf {\bibinfo {volume} {1}},\ \bibinfo
  {pages} {385} (\bibinfo {year} {2010})}\BibitemShut {NoStop}%
\bibitem [{\citenamefont {Shirai}\ \emph {et~al.}(2017)\citenamefont {Shirai},
  \citenamefont {Datta},\ and\ \citenamefont {Oshita}}]{Shirai2017}%
  \BibitemOpen
  \bibfield  {author} {\bibinfo {author} {\bibfnamefont {H.}~\bibnamefont
  {Shirai}}, \bibinfo {author} {\bibfnamefont {A.~K.}\ \bibnamefont {Datta}},\
  and\ \bibinfo {author} {\bibfnamefont {S.}~\bibnamefont {Oshita}},\
  }\bibfield  {title} {\bibinfo {title} {Penetration of aerobic bacteria into
  meat: A mechanistic understanding},\ }\href
  {https://doi.org/https://doi.org/10.1016/j.jfoodeng.2016.10.012} {\bibfield
  {journal} {\bibinfo  {journal} {Journal of Food Engineering}\ }\textbf
  {\bibinfo {volume} {196}},\ \bibinfo {pages} {193 } (\bibinfo {year}
  {2017})}\BibitemShut {NoStop}%
\bibitem [{\citenamefont {Hou}\ \emph {et~al.}(2000)\citenamefont {Hou},
  \citenamefont {Larsen},\ and\ \citenamefont {Boudko}}]{Hou2000}%
  \BibitemOpen
  \bibfield  {author} {\bibinfo {author} {\bibfnamefont {S.}~\bibnamefont
  {Hou}}, \bibinfo {author} {\bibfnamefont {R.}~\bibnamefont {Larsen}},\ and\
  \bibinfo {author} {\bibfnamefont {D.~e.~a.}\ \bibnamefont {Boudko}},\
  }\bibfield  {title} {\bibinfo {title} {Myoglobin-like aerotaxis transducers
  in archaea and bacteria},\ }\href {https://doi.org/10.1038/35000570}
  {\bibfield  {journal} {\bibinfo  {journal} {Nature}\ }\textbf {\bibinfo
  {volume} {403}},\ \bibinfo {pages} {540–544} (\bibinfo {year}
  {2000})}\BibitemShut {NoStop}%
\bibitem [{\citenamefont {Stock}(1997)}]{Stock1997}%
  \BibitemOpen
  \bibfield  {author} {\bibinfo {author} {\bibfnamefont {A.~M.}\ \bibnamefont
  {Stock}},\ }\bibfield  {title} {\bibinfo {title} {Energy sensors for
  aerotaxis in \emph{Escherichia coli}: Something old, something new},\ }\href
  {https://doi.org/10.1073/pnas.94.20.10487} {\bibfield  {journal} {\bibinfo
  {journal} {Proceedings of the National Academy of Sciences}\ }\textbf
  {\bibinfo {volume} {94}},\ \bibinfo {pages} {10487} (\bibinfo {year}
  {1997})}\BibitemShut {NoStop}%
\bibitem [{\citenamefont {Rebbapragada}\ \emph {et~al.}(1997)\citenamefont
  {Rebbapragada}, \citenamefont {Johnson}, \citenamefont {Harding},
  \citenamefont {Zuccarelli}, \citenamefont {Fletcher}, \citenamefont
  {Zhulin},\ and\ \citenamefont {Taylor}}]{rebbapragada1997aer}%
  \BibitemOpen
  \bibfield  {author} {\bibinfo {author} {\bibfnamefont {A.}~\bibnamefont
  {Rebbapragada}}, \bibinfo {author} {\bibfnamefont {M.~S.}\ \bibnamefont
  {Johnson}}, \bibinfo {author} {\bibfnamefont {G.~P.}\ \bibnamefont
  {Harding}}, \bibinfo {author} {\bibfnamefont {A.~J.}\ \bibnamefont
  {Zuccarelli}}, \bibinfo {author} {\bibfnamefont {H.~M.}\ \bibnamefont
  {Fletcher}}, \bibinfo {author} {\bibfnamefont {I.~B.}\ \bibnamefont
  {Zhulin}},\ and\ \bibinfo {author} {\bibfnamefont {B.~L.}\ \bibnamefont
  {Taylor}},\ }\bibfield  {title} {\bibinfo {title} {The aer protein and the
  serine chemoreceptor tsr independently sense intracellular energy levels and
  transduce oxygen, redox, and energy signals for \emph{Escherichia coli}
  behavior},\ }\href@noop {} {\bibfield  {journal} {\bibinfo  {journal}
  {Proceedings of the National Academy of Sciences}\ }\textbf {\bibinfo
  {volume} {94}},\ \bibinfo {pages} {10541} (\bibinfo {year}
  {1997})}\BibitemShut {NoStop}%
\bibitem [{\citenamefont {Bibikov}\ \emph {et~al.}(1997)\citenamefont
  {Bibikov}, \citenamefont {Biran}, \citenamefont {Rudd},\ and\ \citenamefont
  {Parkinson}}]{bibikov1997signal}%
  \BibitemOpen
  \bibfield  {author} {\bibinfo {author} {\bibfnamefont {S.~I.}\ \bibnamefont
  {Bibikov}}, \bibinfo {author} {\bibfnamefont {R.}~\bibnamefont {Biran}},
  \bibinfo {author} {\bibfnamefont {K.~E.}\ \bibnamefont {Rudd}},\ and\
  \bibinfo {author} {\bibfnamefont {J.~S.}\ \bibnamefont {Parkinson}},\
  }\bibfield  {title} {\bibinfo {title} {A signal transducer for aerotaxis in
  \emph{Escherichia coli}.},\ }\href@noop {} {\bibfield  {journal} {\bibinfo
  {journal} {Journal of bacteriology}\ }\textbf {\bibinfo {volume} {179}},\
  \bibinfo {pages} {4075} (\bibinfo {year} {1997})}\BibitemShut {NoStop}%
\bibitem [{\citenamefont {Yu}\ \emph {et~al.}(2002)\citenamefont {Yu},
  \citenamefont {Saw}, \citenamefont {Hou}, \citenamefont {Larsen},
  \citenamefont {Watts}, \citenamefont {Johnson}, \citenamefont {Zimmer},
  \citenamefont {Ordal}, \citenamefont {Taylor},\ and\ \citenamefont
  {Alam}}]{yu_aerotactic_2002}%
  \BibitemOpen
  \bibfield  {author} {\bibinfo {author} {\bibfnamefont {H.~S.}\ \bibnamefont
  {Yu}}, \bibinfo {author} {\bibfnamefont {J.~H.}\ \bibnamefont {Saw}},
  \bibinfo {author} {\bibfnamefont {S.}~\bibnamefont {Hou}}, \bibinfo {author}
  {\bibfnamefont {R.~W.}\ \bibnamefont {Larsen}}, \bibinfo {author}
  {\bibfnamefont {K.~J.}\ \bibnamefont {Watts}}, \bibinfo {author}
  {\bibfnamefont {M.~S.}\ \bibnamefont {Johnson}}, \bibinfo {author}
  {\bibfnamefont {M.~A.}\ \bibnamefont {Zimmer}}, \bibinfo {author}
  {\bibfnamefont {G.~W.}\ \bibnamefont {Ordal}}, \bibinfo {author}
  {\bibfnamefont {B.~L.}\ \bibnamefont {Taylor}},\ and\ \bibinfo {author}
  {\bibfnamefont {M.}~\bibnamefont {Alam}},\ }\bibfield  {title} {\bibinfo
  {title} {Aerotactic responses in bacteria to photoreleased oxygen},\ }\href
  {https://doi.org/10.1111/j.1574-6968.2002.tb11481.x} {\bibfield  {journal}
  {\bibinfo  {journal} {FEMS Microbiology Letters}\ }\textbf {\bibinfo {volume}
  {217}},\ \bibinfo {pages} {237} (\bibinfo {year} {2002})}\BibitemShut
  {NoStop}%
\bibitem [{\citenamefont {Sherris}\ \emph {et~al.}(1957)\citenamefont
  {Sherris}, \citenamefont {Preston},\ and\ \citenamefont
  {Shoesmith}}]{sherris_influence_1957}%
  \BibitemOpen
  \bibfield  {author} {\bibinfo {author} {\bibfnamefont {J.~C.}\ \bibnamefont
  {Sherris}}, \bibinfo {author} {\bibfnamefont {N.~W.}\ \bibnamefont
  {Preston}},\ and\ \bibinfo {author} {\bibfnamefont {J.~G.}\ \bibnamefont
  {Shoesmith}},\ }\bibfield  {title} {\bibinfo {title} {The {Influence} of
  {Oxygen} and {Arginine} on the {Motility} of a {Strain} of {Pseudomonas}
  sp.},\ }\href {https://doi.org/10.1099/00221287-16-1-86} {\bibfield
  {journal} {\bibinfo  {journal} {Microbiology,}\ }\textbf {\bibinfo {volume}
  {16}},\ \bibinfo {pages} {86} (\bibinfo {year} {1957})}\BibitemShut {NoStop}%
\bibitem [{\citenamefont {Saragosti}\ \emph {et~al.}(2011)\citenamefont
  {Saragosti}, \citenamefont {Calvez}, \citenamefont {Bournaveas},
  \citenamefont {Perthame}, \citenamefont {Buguin},\ and\ \citenamefont
  {Silberzan}}]{saragosti2011directional}%
  \BibitemOpen
  \bibfield  {author} {\bibinfo {author} {\bibfnamefont {J.}~\bibnamefont
  {Saragosti}}, \bibinfo {author} {\bibfnamefont {V.}~\bibnamefont {Calvez}},
  \bibinfo {author} {\bibfnamefont {N.}~\bibnamefont {Bournaveas}}, \bibinfo
  {author} {\bibfnamefont {B.}~\bibnamefont {Perthame}}, \bibinfo {author}
  {\bibfnamefont {A.}~\bibnamefont {Buguin}},\ and\ \bibinfo {author}
  {\bibfnamefont {P.}~\bibnamefont {Silberzan}},\ }\bibfield  {title} {\bibinfo
  {title} {Directional persistence of chemotactic bacteria in a traveling
  concentration wave},\ }\href@noop {} {\bibfield  {journal} {\bibinfo
  {journal} {Proceedings of the National Academy of Sciences}\ }\textbf
  {\bibinfo {volume} {108}},\ \bibinfo {pages} {16235} (\bibinfo {year}
  {2011})}\BibitemShut {NoStop}%
\bibitem [{\citenamefont {Cremer}\ \emph {et~al.}(2019)\citenamefont {Cremer},
  \citenamefont {Honda}, \citenamefont {Tang}, \citenamefont {Wong-Ng},
  \citenamefont {Vergassola},\ and\ \citenamefont
  {Hwa}}]{cremer2019chemotaxis}%
  \BibitemOpen
  \bibfield  {author} {\bibinfo {author} {\bibfnamefont {J.}~\bibnamefont
  {Cremer}}, \bibinfo {author} {\bibfnamefont {T.}~\bibnamefont {Honda}},
  \bibinfo {author} {\bibfnamefont {Y.}~\bibnamefont {Tang}}, \bibinfo {author}
  {\bibfnamefont {J.}~\bibnamefont {Wong-Ng}}, \bibinfo {author} {\bibfnamefont
  {M.}~\bibnamefont {Vergassola}},\ and\ \bibinfo {author} {\bibfnamefont
  {T.}~\bibnamefont {Hwa}},\ }\bibfield  {title} {\bibinfo {title} {Chemotaxis
  as a navigation strategy to boost range expansion},\ }\href@noop {}
  {\bibfield  {journal} {\bibinfo  {journal} {Nature}\ }\textbf {\bibinfo
  {volume} {575}},\ \bibinfo {pages} {658} (\bibinfo {year}
  {2019})}\BibitemShut {NoStop}%
\bibitem [{\citenamefont {Keller}\ and\ \citenamefont
  {Segel}(1971)}]{keller1971traveling}%
  \BibitemOpen
  \bibfield  {author} {\bibinfo {author} {\bibfnamefont {E.~F.}\ \bibnamefont
  {Keller}}\ and\ \bibinfo {author} {\bibfnamefont {L.~A.}\ \bibnamefont
  {Segel}},\ }\bibfield  {title} {\bibinfo {title} {Traveling bands of
  chemotactic bacteria: a theoretical analysis},\ }\href@noop {} {\bibfield
  {journal} {\bibinfo  {journal} {Journal of theoretical biology}\ }\textbf
  {\bibinfo {volume} {30}},\ \bibinfo {pages} {235} (\bibinfo {year}
  {1971})}\BibitemShut {NoStop}%
\bibitem [{\citenamefont {Rivero}\ \emph {et~al.}(1989)\citenamefont {Rivero},
  \citenamefont {Tranquillo}, \citenamefont {Buettner},\ and\ \citenamefont
  {Lauffenburger}}]{rivero1989transport}%
  \BibitemOpen
  \bibfield  {author} {\bibinfo {author} {\bibfnamefont {M.~A.}\ \bibnamefont
  {Rivero}}, \bibinfo {author} {\bibfnamefont {R.~T.}\ \bibnamefont
  {Tranquillo}}, \bibinfo {author} {\bibfnamefont {H.~M.}\ \bibnamefont
  {Buettner}},\ and\ \bibinfo {author} {\bibfnamefont {D.~A.}\ \bibnamefont
  {Lauffenburger}},\ }\bibfield  {title} {\bibinfo {title} {Transport models
  for chemotactic cell populations based on individual cell behavior},\
  }\href@noop {} {\bibfield  {journal} {\bibinfo  {journal} {Chemical
  engineering science}\ }\textbf {\bibinfo {volume} {44}},\ \bibinfo {pages}
  {2881} (\bibinfo {year} {1989})}\BibitemShut {NoStop}%
\bibitem [{\citenamefont {Lapidus}\ and\ \citenamefont
  {Schiller}(1976)}]{lapidus1976model}%
  \BibitemOpen
  \bibfield  {author} {\bibinfo {author} {\bibfnamefont {I.~R.}\ \bibnamefont
  {Lapidus}}\ and\ \bibinfo {author} {\bibfnamefont {R.}~\bibnamefont
  {Schiller}},\ }\bibfield  {title} {\bibinfo {title} {Model for the
  chemotactic response of a bacterial population},\ }\href@noop {} {\bibfield
  {journal} {\bibinfo  {journal} {Biophysical journal}\ }\textbf {\bibinfo
  {volume} {16}},\ \bibinfo {pages} {779} (\bibinfo {year} {1976})}\BibitemShut
  {NoStop}%
\bibitem [{\citenamefont {Shioi}\ \emph {et~al.}(1987)\citenamefont {Shioi},
  \citenamefont {Dang},\ and\ \citenamefont {Taylor}}]{shioi1987oxygen}%
  \BibitemOpen
  \bibfield  {author} {\bibinfo {author} {\bibfnamefont {J.}~\bibnamefont
  {Shioi}}, \bibinfo {author} {\bibfnamefont {C.}~\bibnamefont {Dang}},\ and\
  \bibinfo {author} {\bibfnamefont {B.}~\bibnamefont {Taylor}},\ }\bibfield
  {title} {\bibinfo {title} {Oxygen as attractant and repellent in bacterial
  chemotaxis},\ }\href@noop {} {\bibfield  {journal} {\bibinfo  {journal}
  {Journal of Bacteriology}\ }\textbf {\bibinfo {volume} {169}},\ \bibinfo
  {pages} {3118} (\bibinfo {year} {1987})}\BibitemShut {NoStop}%
\bibitem [{\citenamefont {Kalinin}\ \emph {et~al.}(2009)\citenamefont
  {Kalinin}, \citenamefont {Jiang}, \citenamefont {Tu},\ and\ \citenamefont
  {Wu}}]{kalinin2009}%
  \BibitemOpen
  \bibfield  {author} {\bibinfo {author} {\bibfnamefont {Y.~V.}\ \bibnamefont
  {Kalinin}}, \bibinfo {author} {\bibfnamefont {L.}~\bibnamefont {Jiang}},
  \bibinfo {author} {\bibfnamefont {Y.}~\bibnamefont {Tu}},\ and\ \bibinfo
  {author} {\bibfnamefont {M.}~\bibnamefont {Wu}},\ }\bibfield  {title}
  {\bibinfo {title} {Logarithmic sensing in escherichia coli bacterial
  chemotaxis},\ }\href
  {https://doi.org/https://doi.org/10.1016/j.bpj.2008.10.027} {\bibfield
  {journal} {\bibinfo  {journal} {Biophysical Journal}\ }\textbf {\bibinfo
  {volume} {96}},\ \bibinfo {pages} {2439 } (\bibinfo {year}
  {2009})}\BibitemShut {NoStop}%
\bibitem [{\citenamefont {Zhuang}\ \emph {et~al.}(2014)\citenamefont {Zhuang},
  \citenamefont {Wei}, \citenamefont {Carlsen}, \citenamefont {Edwards},
  \citenamefont {Marculescu}, \citenamefont {Bogdan},\ and\ \citenamefont
  {Sitti}}]{zhuang2014analytical}%
  \BibitemOpen
  \bibfield  {author} {\bibinfo {author} {\bibfnamefont {J.}~\bibnamefont
  {Zhuang}}, \bibinfo {author} {\bibfnamefont {G.}~\bibnamefont {Wei}},
  \bibinfo {author} {\bibfnamefont {R.~W.}\ \bibnamefont {Carlsen}}, \bibinfo
  {author} {\bibfnamefont {M.~R.}\ \bibnamefont {Edwards}}, \bibinfo {author}
  {\bibfnamefont {R.}~\bibnamefont {Marculescu}}, \bibinfo {author}
  {\bibfnamefont {P.}~\bibnamefont {Bogdan}},\ and\ \bibinfo {author}
  {\bibfnamefont {M.}~\bibnamefont {Sitti}},\ }\bibfield  {title} {\bibinfo
  {title} {Analytical modeling and experimental characterization of chemotaxis
  in serratia marcescens},\ }\href@noop {} {\bibfield  {journal} {\bibinfo
  {journal} {Physical Review E}\ }\textbf {\bibinfo {volume} {89}},\ \bibinfo
  {pages} {052704} (\bibinfo {year} {2014})}\BibitemShut {NoStop}%
\bibitem [{\citenamefont {Kirkegaard}\ \emph {et~al.}(2016)\citenamefont
  {Kirkegaard}, \citenamefont {Bouillant}, \citenamefont {Marron},
  \citenamefont {Leptos},\ and\ \citenamefont {R.E.}}]{kirkegaard_2016}%
  \BibitemOpen
  \bibfield  {author} {\bibinfo {author} {\bibfnamefont {J.}~\bibnamefont
  {Kirkegaard}}, \bibinfo {author} {\bibfnamefont {A.}~\bibnamefont
  {Bouillant}}, \bibinfo {author} {\bibfnamefont {A.}~\bibnamefont {Marron}},
  \bibinfo {author} {\bibfnamefont {K.}~\bibnamefont {Leptos}},\ and\ \bibinfo
  {author} {\bibfnamefont {G.}~\bibnamefont {R.E.}},\ }\bibfield  {title}
  {\bibinfo {title} {Aerotaxis in the closest relatives of animals},\
  }\bibfield  {journal} {\bibinfo  {journal} {eLife}\ }\textbf {\bibinfo
  {volume} {5}},\ \href {https://doi.org/10.7554/eLife.18109}
  {10.7554/eLife.18109} (\bibinfo {year} {2016})\BibitemShut {NoStop}%
\bibitem [{\citenamefont {Ahmed}\ \emph {et~al.}(2010)\citenamefont {Ahmed},
  \citenamefont {Shimizu},\ and\ \citenamefont
  {Stocker}}]{ahmed2010microfluidics}%
  \BibitemOpen
  \bibfield  {author} {\bibinfo {author} {\bibfnamefont {T.}~\bibnamefont
  {Ahmed}}, \bibinfo {author} {\bibfnamefont {T.~S.}\ \bibnamefont {Shimizu}},\
  and\ \bibinfo {author} {\bibfnamefont {R.}~\bibnamefont {Stocker}},\
  }\bibfield  {title} {\bibinfo {title} {Microfluidics for bacterial
  chemotaxis},\ }\href@noop {} {\bibfield  {journal} {\bibinfo  {journal}
  {Integrative Biology}\ }\textbf {\bibinfo {volume} {2}},\ \bibinfo {pages}
  {604} (\bibinfo {year} {2010})}\BibitemShut {NoStop}%
\bibitem [{\citenamefont {Bhattacharjee}\ \emph {et~al.}(2021)\citenamefont
  {Bhattacharjee}, \citenamefont {Amchin}, \citenamefont {Ott}, \citenamefont
  {Kratz},\ and\ \citenamefont {Datta}}]{bhattacharjee2021chemotactic}%
  \BibitemOpen
  \bibfield  {author} {\bibinfo {author} {\bibfnamefont {T.}~\bibnamefont
  {Bhattacharjee}}, \bibinfo {author} {\bibfnamefont {D.~B.}\ \bibnamefont
  {Amchin}}, \bibinfo {author} {\bibfnamefont {J.~A.}\ \bibnamefont {Ott}},
  \bibinfo {author} {\bibfnamefont {F.}~\bibnamefont {Kratz}},\ and\ \bibinfo
  {author} {\bibfnamefont {S.~S.}\ \bibnamefont {Datta}},\ }\bibfield  {title}
  {\bibinfo {title} {Chemotactic migration of bacteria in porous media},\
  }\href@noop {} {\bibfield  {journal} {\bibinfo  {journal} {Biophysical
  Journal}\ } (\bibinfo {year} {2021})}\BibitemShut {NoStop}%
\bibitem [{\citenamefont {Alert}\ \emph {et~al.}(2022)\citenamefont {Alert},
  \citenamefont {Mart{\'\i}nez-Calvo},\ and\ \citenamefont
  {Datta}}]{alert2022cellular}%
  \BibitemOpen
  \bibfield  {author} {\bibinfo {author} {\bibfnamefont {R.}~\bibnamefont
  {Alert}}, \bibinfo {author} {\bibfnamefont {A.}~\bibnamefont
  {Mart{\'\i}nez-Calvo}},\ and\ \bibinfo {author} {\bibfnamefont {S.~S.}\
  \bibnamefont {Datta}},\ }\bibfield  {title} {\bibinfo {title} {Cellular
  sensing governs the stability of chemotactic fronts},\ }\href@noop {}
  {\bibfield  {journal} {\bibinfo  {journal} {Physical Review Letters}\
  }\textbf {\bibinfo {volume} {128}},\ \bibinfo {pages} {148101} (\bibinfo
  {year} {2022})}\BibitemShut {NoStop}%
\bibitem [{\citenamefont {Douarche}\ \emph {et~al.}(2009)\citenamefont
  {Douarche}, \citenamefont {Buguin}, \citenamefont {Salman},\ and\
  \citenamefont {Libchaber}}]{douarche2009coli}%
  \BibitemOpen
  \bibfield  {author} {\bibinfo {author} {\bibfnamefont {C.}~\bibnamefont
  {Douarche}}, \bibinfo {author} {\bibfnamefont {A.}~\bibnamefont {Buguin}},
  \bibinfo {author} {\bibfnamefont {H.}~\bibnamefont {Salman}},\ and\ \bibinfo
  {author} {\bibfnamefont {A.}~\bibnamefont {Libchaber}},\ }\bibfield  {title}
  {\bibinfo {title} {\emph{E. coli} and oxygen: a motility transition},\
  }\href@noop {} {\bibfield  {journal} {\bibinfo  {journal} {Physical review
  letters}\ }\textbf {\bibinfo {volume} {102}},\ \bibinfo {pages} {198101}
  (\bibinfo {year} {2009})}\BibitemShut {NoStop}%
\bibitem [{\citenamefont {Tinevez}\ \emph {et~al.}(2017)\citenamefont
  {Tinevez}, \citenamefont {Perry}, \citenamefont {Schindelin}, \citenamefont
  {Hoopes}, \citenamefont {Reynolds}, \citenamefont {Laplantine}, \citenamefont
  {Bednarek}, \citenamefont {Shorte},\ and\ \citenamefont
  {Eliceiri}}]{tinevez2017trackmate}%
  \BibitemOpen
  \bibfield  {author} {\bibinfo {author} {\bibfnamefont {J.-Y.}\ \bibnamefont
  {Tinevez}}, \bibinfo {author} {\bibfnamefont {N.}~\bibnamefont {Perry}},
  \bibinfo {author} {\bibfnamefont {J.}~\bibnamefont {Schindelin}}, \bibinfo
  {author} {\bibfnamefont {G.~M.}\ \bibnamefont {Hoopes}}, \bibinfo {author}
  {\bibfnamefont {G.~D.}\ \bibnamefont {Reynolds}}, \bibinfo {author}
  {\bibfnamefont {E.}~\bibnamefont {Laplantine}}, \bibinfo {author}
  {\bibfnamefont {S.~Y.}\ \bibnamefont {Bednarek}}, \bibinfo {author}
  {\bibfnamefont {S.~L.}\ \bibnamefont {Shorte}},\ and\ \bibinfo {author}
  {\bibfnamefont {K.~W.}\ \bibnamefont {Eliceiri}},\ }\bibfield  {title}
  {\bibinfo {title} {Trackmate: An open and extensible platform for
  single-particle tracking},\ }\href@noop {} {\bibfield  {journal} {\bibinfo
  {journal} {Methods}\ }\textbf {\bibinfo {volume} {115}},\ \bibinfo {pages}
  {80} (\bibinfo {year} {2017})}\BibitemShut {NoStop}%
\bibitem [{\citenamefont {Bouvard}(2022)}]{bouvard2022thesis}%
  \BibitemOpen
  \bibfield  {author} {\bibinfo {author} {\bibfnamefont {J.}~\bibnamefont
  {Bouvard}},\ }\emph {\bibinfo {title} {{Dynamics of bacterial suspensions,
  from aerotaxis to cluster formation}}},\ \href
  {https://tel.archives-ouvertes.fr/tel-03680098} {\bibinfo {type} {Theses}},\
  \bibinfo  {school} {{Universit{\'e} Paris-Saclay}} (\bibinfo {year}
  {2022})\BibitemShut {NoStop}%
\bibitem [{\citenamefont {Berg}(1993)}]{berg1993random}%
  \BibitemOpen
  \bibfield  {author} {\bibinfo {author} {\bibfnamefont {H.~C.}\ \bibnamefont
  {Berg}},\ }\href@noop {} {\emph {\bibinfo {title} {Random Walks in
  Biology}}}\ (\bibinfo  {publisher} {Princeton University Press},\ \bibinfo
  {year} {1993})\BibitemShut {NoStop}%
\bibitem [{\citenamefont {Barbara}\ and\ \citenamefont
  {Mitchell}(2003)}]{barbara2003bacterial}%
  \BibitemOpen
  \bibfield  {author} {\bibinfo {author} {\bibfnamefont {G.~M.}\ \bibnamefont
  {Barbara}}\ and\ \bibinfo {author} {\bibfnamefont {J.~G.}\ \bibnamefont
  {Mitchell}},\ }\bibfield  {title} {\bibinfo {title} {Bacterial tracking of
  motile algae},\ }\href@noop {} {\bibfield  {journal} {\bibinfo  {journal}
  {FEMS microbiology ecology}\ }\textbf {\bibinfo {volume} {44}},\ \bibinfo
  {pages} {79} (\bibinfo {year} {2003})}\BibitemShut {NoStop}%
\bibitem [{\citenamefont {Taktikos}\ \emph {et~al.}(2013)\citenamefont
  {Taktikos}, \citenamefont {Stark},\ and\ \citenamefont
  {Zaburdaev}}]{taktikos2013motility}%
  \BibitemOpen
  \bibfield  {author} {\bibinfo {author} {\bibfnamefont {J.}~\bibnamefont
  {Taktikos}}, \bibinfo {author} {\bibfnamefont {H.}~\bibnamefont {Stark}},\
  and\ \bibinfo {author} {\bibfnamefont {V.}~\bibnamefont {Zaburdaev}},\
  }\bibfield  {title} {\bibinfo {title} {How the motility pattern of bacteria
  affects their dispersal and chemotaxis},\ }\href@noop {} {\bibfield
  {journal} {\bibinfo  {journal} {PloS one}\ }\textbf {\bibinfo {volume} {8}},\
  \bibinfo {pages} {e81936} (\bibinfo {year} {2013})}\BibitemShut {NoStop}%
\bibitem [{\citenamefont {Grognot}\ and\ \citenamefont
  {Taute}(2021{\natexlab{a}})}]{grognot2021multiscale}%
  \BibitemOpen
  \bibfield  {author} {\bibinfo {author} {\bibfnamefont {M.}~\bibnamefont
  {Grognot}}\ and\ \bibinfo {author} {\bibfnamefont {K.~M.}\ \bibnamefont
  {Taute}},\ }\bibfield  {title} {\bibinfo {title} {A multiscale 3d chemotaxis
  assay reveals bacterial navigation mechanisms},\ }\href@noop {} {\bibfield
  {journal} {\bibinfo  {journal} {Communications biology}\ }\textbf {\bibinfo
  {volume} {4}},\ \bibinfo {pages} {1} (\bibinfo {year}
  {2021}{\natexlab{a}})}\BibitemShut {NoStop}%
\bibitem [{\citenamefont {Grognot}\ and\ \citenamefont
  {Taute}(2021{\natexlab{b}})}]{grognot_2021_propellers}%
  \BibitemOpen
  \bibfield  {author} {\bibinfo {author} {\bibfnamefont {M.}~\bibnamefont
  {Grognot}}\ and\ \bibinfo {author} {\bibfnamefont {K.}~\bibnamefont
  {Taute}},\ }\bibfield  {title} {\bibinfo {title} {More than propellers: how
  flagella shape bacterial motility behaviors},\ }\bibfield  {journal}
  {\bibinfo  {journal} {Current Opinion in Microbiology}\ }\textbf {\bibinfo
  {volume} {61}},\ \href {https://doi.org/10.1016/j.mib.2021.02.005}
  {10.1016/j.mib.2021.02.005} (\bibinfo {year}
  {2021}{\natexlab{b}})\BibitemShut {NoStop}%
\bibitem [{\citenamefont {Gerritsen}\ \emph {et~al.}(1997)\citenamefont
  {Gerritsen}, \citenamefont {Sanders}, \citenamefont {Draaijer}, \citenamefont
  {Ince},\ and\ \citenamefont {Levine}}]{gerritsen1997fluorescence}%
  \BibitemOpen
  \bibfield  {author} {\bibinfo {author} {\bibfnamefont {H.~C.}\ \bibnamefont
  {Gerritsen}}, \bibinfo {author} {\bibfnamefont {R.}~\bibnamefont {Sanders}},
  \bibinfo {author} {\bibfnamefont {A.}~\bibnamefont {Draaijer}}, \bibinfo
  {author} {\bibfnamefont {C.}~\bibnamefont {Ince}},\ and\ \bibinfo {author}
  {\bibfnamefont {Y.}~\bibnamefont {Levine}},\ }\bibfield  {title} {\bibinfo
  {title} {Fluorescence lifetime imaging of oxygen in living cells},\
  }\href@noop {} {\bibfield  {journal} {\bibinfo  {journal} {Journal of
  Fluorescence}\ }\textbf {\bibinfo {volume} {7}},\ \bibinfo {pages} {11}
  (\bibinfo {year} {1997})}\BibitemShut {NoStop}%
\bibitem [{\citenamefont {Sud}\ \emph {et~al.}(2006)\citenamefont {Sud},
  \citenamefont {Zhong}, \citenamefont {Beer},\ and\ \citenamefont
  {Mycek}}]{sud2006time}%
  \BibitemOpen
  \bibfield  {author} {\bibinfo {author} {\bibfnamefont {D.}~\bibnamefont
  {Sud}}, \bibinfo {author} {\bibfnamefont {W.}~\bibnamefont {Zhong}}, \bibinfo
  {author} {\bibfnamefont {D.~G.}\ \bibnamefont {Beer}},\ and\ \bibinfo
  {author} {\bibfnamefont {M.-A.}\ \bibnamefont {Mycek}},\ }\bibfield  {title}
  {\bibinfo {title} {Time-resolved optical imaging provides a molecular
  snapshot of altered metabolic function in living human cancer cell models},\
  }\href@noop {} {\bibfield  {journal} {\bibinfo  {journal} {Optics Express}\
  }\textbf {\bibinfo {volume} {14}},\ \bibinfo {pages} {4412} (\bibinfo {year}
  {2006})}\BibitemShut {NoStop}%
\bibitem [{\citenamefont {Polinkovsky}\ \emph {et~al.}(2009)\citenamefont
  {Polinkovsky}, \citenamefont {Gutierrez}, \citenamefont {Levchenko},\ and\
  \citenamefont {Groisman}}]{polinkovsky2009fine}%
  \BibitemOpen
  \bibfield  {author} {\bibinfo {author} {\bibfnamefont {M.}~\bibnamefont
  {Polinkovsky}}, \bibinfo {author} {\bibfnamefont {E.}~\bibnamefont
  {Gutierrez}}, \bibinfo {author} {\bibfnamefont {A.}~\bibnamefont
  {Levchenko}},\ and\ \bibinfo {author} {\bibfnamefont {A.}~\bibnamefont
  {Groisman}},\ }\bibfield  {title} {\bibinfo {title} {Fine temporal control of
  the medium gas content and acidity and on-chip generation of series of oxygen
  concentrations for cell cultures},\ }\href@noop {} {\bibfield  {journal}
  {\bibinfo  {journal} {Lab on a Chip}\ }\textbf {\bibinfo {volume} {9}},\
  \bibinfo {pages} {1073} (\bibinfo {year} {2009})}\BibitemShut {NoStop}%
\bibitem [{\citenamefont {Kim}\ \emph {et~al.}(2016)\citenamefont {Kim},
  \citenamefont {Chu}, \citenamefont {Jusuf}, \citenamefont {Kuo},
  \citenamefont {TerAvest}, \citenamefont {Angenent},\ and\ \citenamefont
  {Wu}}]{kim2016oxygen}%
  \BibitemOpen
  \bibfield  {author} {\bibinfo {author} {\bibfnamefont {B.~J.}\ \bibnamefont
  {Kim}}, \bibinfo {author} {\bibfnamefont {I.}~\bibnamefont {Chu}}, \bibinfo
  {author} {\bibfnamefont {S.}~\bibnamefont {Jusuf}}, \bibinfo {author}
  {\bibfnamefont {T.}~\bibnamefont {Kuo}}, \bibinfo {author} {\bibfnamefont
  {M.~A.}\ \bibnamefont {TerAvest}}, \bibinfo {author} {\bibfnamefont {L.~T.}\
  \bibnamefont {Angenent}},\ and\ \bibinfo {author} {\bibfnamefont
  {M.}~\bibnamefont {Wu}},\ }\bibfield  {title} {\bibinfo {title} {Oxygen
  tension and riboflavin gradients cooperatively regulate the migration of
  shewanella oneidensis mr-1 revealed by a hydrogel-based microfluidic
  device},\ }\href@noop {} {\bibfield  {journal} {\bibinfo  {journal}
  {Frontiers in microbiology}\ }\textbf {\bibinfo {volume} {7}},\ \bibinfo
  {pages} {1438} (\bibinfo {year} {2016})}\BibitemShut {NoStop}%
\bibitem [{\citenamefont {Lovely}\ and\ \citenamefont
  {Dahlquist}(1975)}]{lovely1975statistical}%
  \BibitemOpen
  \bibfield  {author} {\bibinfo {author} {\bibfnamefont {P.~S.}\ \bibnamefont
  {Lovely}}\ and\ \bibinfo {author} {\bibfnamefont {F.}~\bibnamefont
  {Dahlquist}},\ }\bibfield  {title} {\bibinfo {title} {Statistical measures of
  bacterial motility and chemotaxis},\ }\href@noop {} {\bibfield  {journal}
  {\bibinfo  {journal} {Journal of theoretical biology}\ }\textbf {\bibinfo
  {volume} {50}},\ \bibinfo {pages} {477} (\bibinfo {year} {1975})}\BibitemShut
  {NoStop}%
\bibitem [{\citenamefont {Lauga}(2020)}]{lauga2020}%
  \BibitemOpen
  \bibfield  {author} {\bibinfo {author} {\bibfnamefont {E.}~\bibnamefont
  {Lauga}},\ }\href@noop {} {\emph {\bibinfo {title} {The Fluid Dynamics of
  Cell Motility}}}\ (\bibinfo  {publisher} {Cambridge University Press},\
  \bibinfo {year} {2020})\BibitemShut {NoStop}%
\end{thebibliography}%

\section{Concentration profiles}
\label{sec:SM_compil_profiles}

The dynamics of the migration band, shown in Fig. 3 at
four selected times for the bacterial concentration OD = 0.2, 
is complemented in Figs. 13, 14 and 15 at all times for the 
three concentrations OD = 0.2, 0.1 and 0.05.

\begin{figure*}[ht]
\includegraphics[trim = 0mm 0mm 0mm 0mm, clip, width=0.8\textwidth, angle=0]{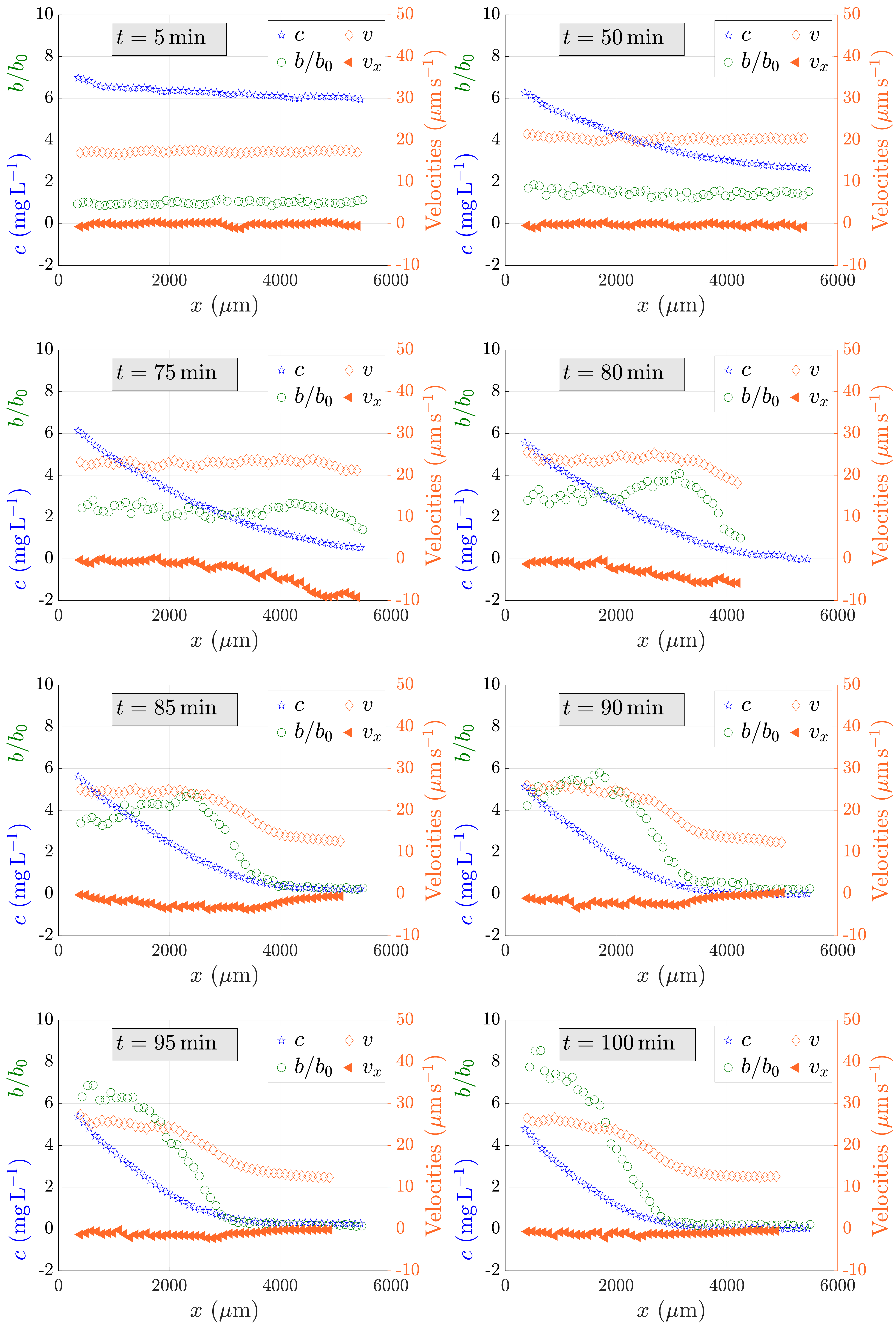}
\caption{Variations of the local density of bacteria $b(x)$ (green circles), oxygen concentration $c(x)$ (blue stars), magnitude of the swimming velocity of bacteria $v(x)$ (empty orange diamonds) and drift velocity along the gradient $v_x(x)$ (filled orange diamonds) as functions of the distance from the oxygen source at four different times during the migration of the bacterial band. The capillary is initially filled with an homogeneous suspension of bacteria at a concentration $\text{OD}=0.05$. Time is counted from the sealing of the capillary.}   
\label{fig:SM_Compil_OD=0.05}
\end{figure*}

\begin{figure*}[ht]
\includegraphics[trim = 0mm 0mm 0mm 0mm, clip, width=0.8\textwidth, angle=0]{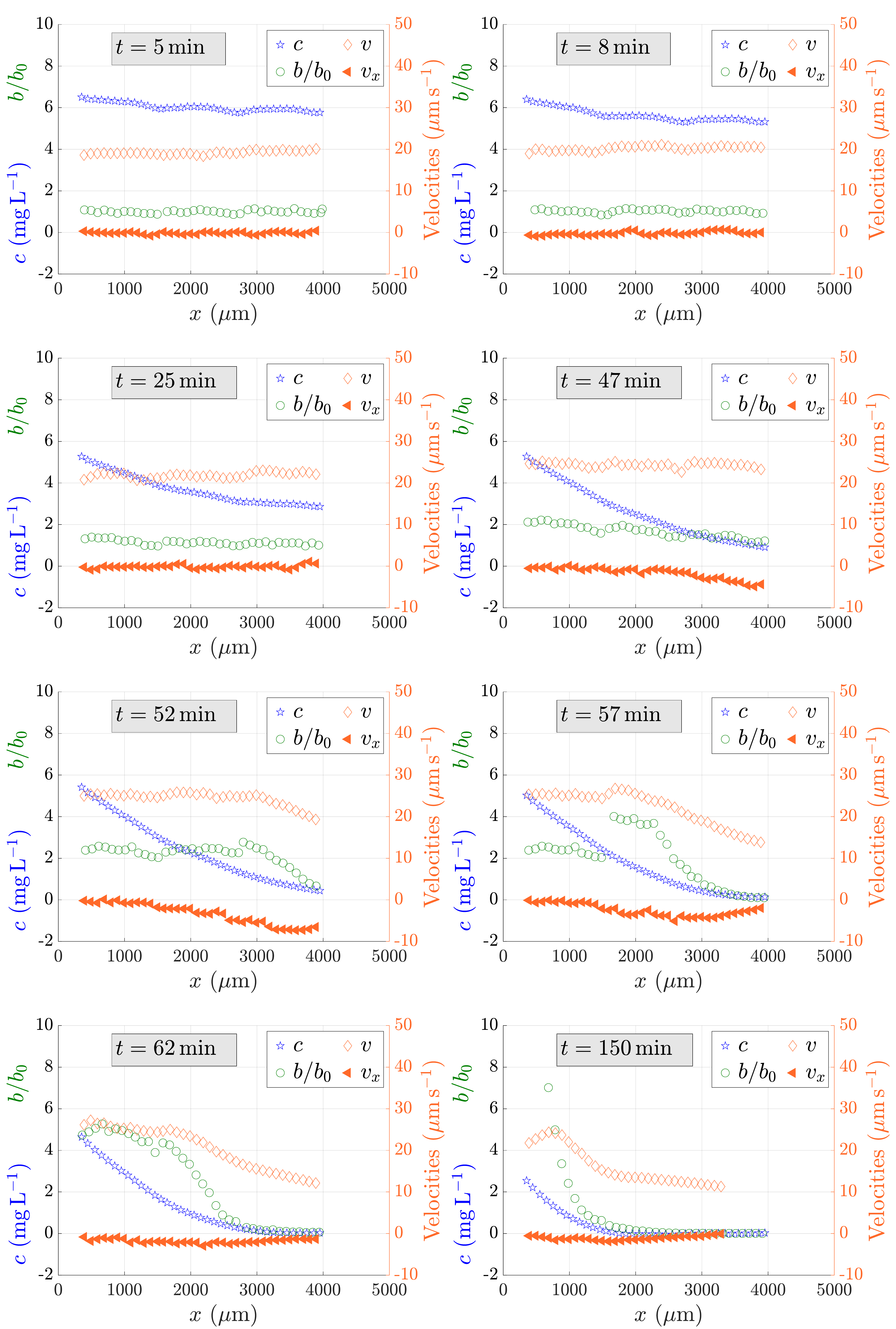}
\caption{Same as described in the caption of Fig.~\ref{fig:SM_Compil_OD=0.05} for $\mathrm{OD}=0.1$.}  
\label{fig:SM_Compil_OD=0.1}
\end{figure*}

\begin{figure*}[ht]
\includegraphics[trim = 0mm 0mm 0mm 0mm, clip, width=0.8\textwidth, angle=0]{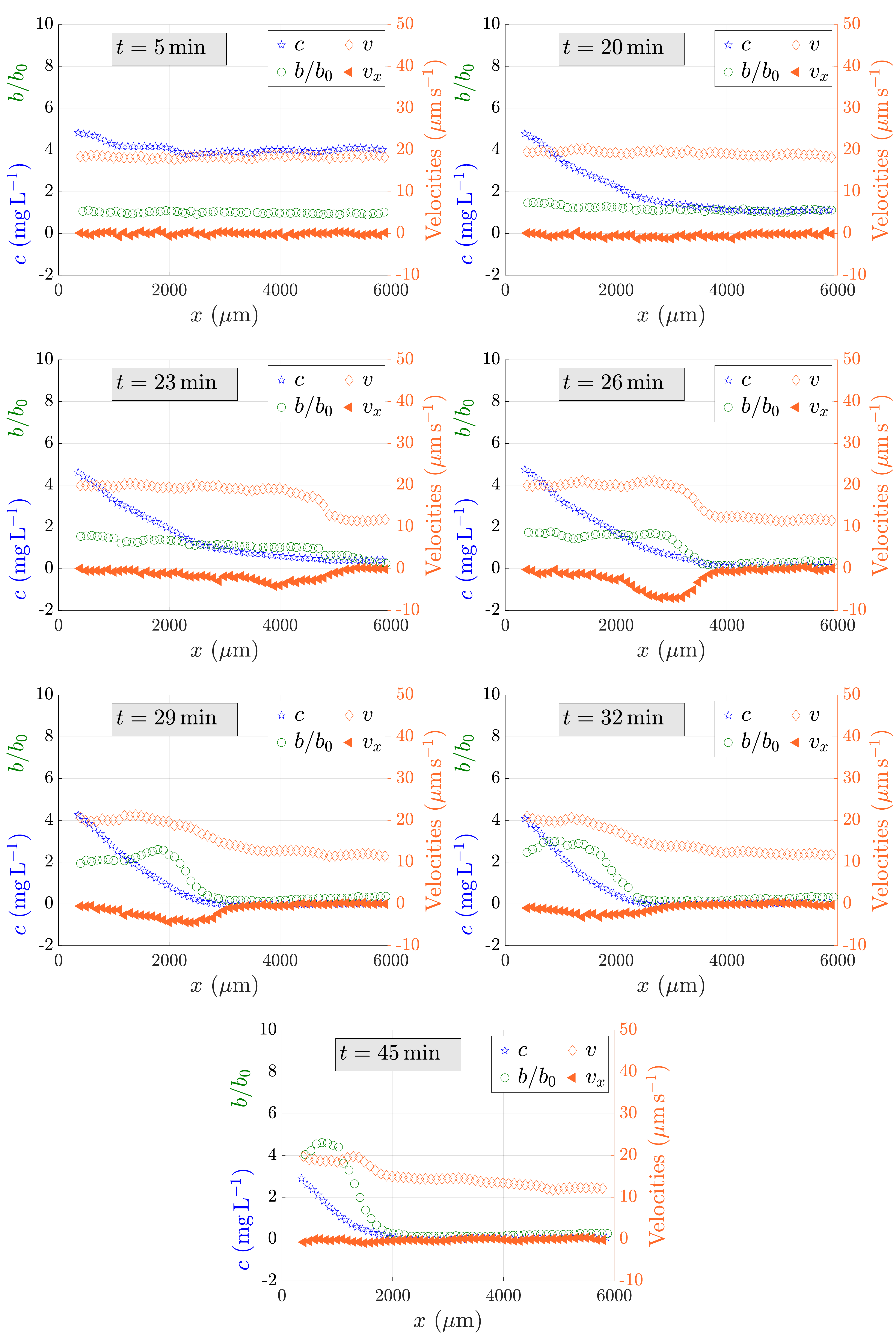}
\caption{Same as described in the caption of  Fig.~\ref{fig:SM_Compil_OD=0.05} for $\mathrm{OD}=0.2$.}    
\label{fig:SM_Compil_OD=0.2}
\end{figure*}

\end{document}